\documentclass{article}

\usepackage{arxiv}

\usepackage[utf8]{inputenc} % allow utf-8 input
\usepackage[T1]{fontenc}    % use 8-bit T1 fonts
\usepackage{hyperref}       % hyperlinks
\usepackage{url}            % simple URL typesetting
\usepackage{booktabs}       % professional-quality tables
\usepackage{amsfonts}       % blackboard math symbols
\usepackage{nicefrac}       % compact symbols for 1/2, etc.
\usepackage{microtype}      % microtypography
\usepackage{graphicx}
\usepackage{xcolor}

\usepackage{doi}

\usepackage{algorithm}
\usepackage{algorithmic}
\usepackage{amsmath}
\usepackage{tabularx}
\usepackage{subcaption}
\usepackage{eqparbox}
\usepackage{comment}

\usepackage[framemethod=TikZ]{mdframed}
\mdfdefinestyle{style1}{
    backgroundcolor=gray!10,
	innerleftmargin=0.3cm,innerrightmargin=0.3cm,
	innertopmargin=0.3cm ,innerbottommargin=0.3cm,
	roundcorner=4pt,linewidth=0.6pt,
	footnoteinside=false}

\title{Automated Black-Box Boundary Value Detection
}

\author{
  Felix Dobslaw \\
  Dept.\ of Computer and System Sciences \\
  Mid Sweden University \\
  Östersund\\
  \texttt{felix.dobslaw@miun.se} \\
   \And
    Robert Feldt, Francisco Gomes de Oliveira Neto \\
  Dept.\ of Computer Science and Engineering \\
  Chalmers, and the University of Gothenburg \\
  Gothenburg\\
  \texttt{robert.feldt@chalmers.se}, \texttt{francisco.gomes@cse.gu.se} \\
}

\begin{document}
\maketitle

\begin{abstract}
The input domain of software systems can typically be divided into sub-domains for which the outputs are similar. To ensure high quality it is critical to test the software on the boundaries between these sub-domains. Consequently, boundary value analysis and testing has been part of the toolbox of software testers for long and is typically taught early to students. However, despite its many argued benefits, boundary value analysis for a given specification or piece of software is typically described in abstract terms which allow for variation in how testers apply it.

Here we propose an automated, black-box boundary value detection method to support software testers in systematic boundary value analysis with consistent results. The method builds on a metric to quantify the level of boundariness of test inputs: the program derivative. By coupling it with search algorithms we find and rank pairs of inputs as good boundary candidates, i.e. inputs close together but with outputs far apart. We implement our AutoBVA approach and evaluate it on a curated dataset of example programs. Our results indicate that even with a simple and generic program derivative variant in combination with broad sampling over the input space, interesting boundary candidates can be identified.
\end{abstract}


\section{Introduction}

Ensuring the quality of software is critical and while much progress has been made to improve formal verification approaches, testing is still the de facto method and a key part of modern software development. A central problem in software testing is how to meaningfully and efficiently cover an input space that is typically very large. A fundamental, simplifying assumption, is that even for such large input spaces there are subsets of inputs, called partitions or sub-domains, that the software will handle in the same or similar way~\cite{goodenough1975toward,richardson1985partition,hamlet1990partition,grindal2005combination}.
Thus, if we can identify such partitions we only need to select a few inputs from each partition to test and ensure they are correctly handled.

While many different approaches to partition testing have been proposed~\cite{goodenough1975toward,richardson1985partition,ostrand1988category,grochtmann1993classification} one that comes naturally to many testers is boundary value analysis (BVA) and testing~\cite{myers1979,white1980domain,clarke1982close}. 
It is based on the intuition that developers are more likely to get things wrong around the boundaries between input partitions, i.e. where there should be a change in how the input is processed and in the output that is produced~\cite{clarke1982close}. By analysing a specification, testers can identify partitions and boundaries between them. They should then select test inputs on either side of these boundaries and thus, ideally, verify both correct behavior in the partitions as well as that the boundary between them is in the expected, correct place~\cite{clarke1982close,BCS2001}.
But note that identifying the boundaries, and thus partitions, is the critical step; a tester can then decide whether to focus only on them or to also sample some non-boundary inputs ``inside'' of each partition.

A problem with partition testing, in general, and boundary value analysis (BVA), in particular, is that there is no clear and objective method for how to identify partitions nor the boundaries between them. Already Myers~\cite{myers1979} pointed out the difficulty of presenting a ``cookbook'' method and that testers need to be creative and adapt to the software being tested. Later work describe BVA and partition testing as relying on either a partition model~\cite{BCS2001}, categories\slash classifications of inputs or environment conditions~\cite{ostrand1988category,grochtmann1993classification}, constraints~\cite{richardson1985partition,ostrand1988category}, or ``checkpoints''~\cite{yin1997automatic} that are all to be derived from the specification. 
But they don't guide this derivation step in detail. Several authors have also pointed to the fact that existing methods don't give enough support to testers~\cite{grindal2005combination} and that BVA is, at its core, a manual and creative process that cannot be automated~\cite{grochtmann1993classification}.
More recent work can be seen as overcoming the problem by proposing equivalence partitioning and boundary-guided testing from formal models~\cite{hubner2019experimental}. However, this assumes that such models are (readily) available or can be derived, and then maintained, without major costs.

One alternative is to view and then provide tooling for using BVA as a white-box testing technique.
Pandita et al~\cite{pandita2010guided} use instrumentation of control flow expressions and dynamic, symbolic execution to generate test cases that increases boundary value coverage~\cite{kosmatov2004boundary}.
However, it is not clear how such boundaries, internal to the code, relate to the boundaries that traditional, black-box BVA would find. Furthermore, it requires instrumentation and advanced tooling which is not always available and might be costly. % for all programming languages or environments.
It should be noted that white-box testing is limited to situations where source code is available; black-box testing approaches do not have this limitation.

Empirical studies on the effectiveness of partition and boundary value testing do not provide a clear picture. While early work claimed that random testing was as or more effective~\cite{hamlet1990partition}, they were later countered by studies showing clear benefits to BVA~\cite{reid1997empirical,yin1997automatic}. A more recent overview of the debate also provided theoretical results on effectiveness and discussed the scalability of random testing in relation to partition testing methods~\cite{arcuri2011random}. Regardless of the relative benefits of the respective techniques, we argue that improving partition testing and boundary value analysis has both practical and scientific value; judging their value will ultimately depend on how applicable and automatic they can be made.

Here we address the core problem of \textit{how to automate black-box boundary value analysis}.
We build on our recent work that proposed a family of metrics to quantify the boundariness of software %skipped: _of pairs_ of software inputs
inputs~\cite{feldtdobslaw2019} and combine it with search and optimization algorithms in order to automatically detect promising boundary candidates. These can then be (interactively) presented to testers and developers to help them explore meaningful boundary values and create corresponding test cases~\cite{dobslaw2020boundary}.
Our search-based, black-box, and automated boundary value identification method does not require manual analysis of a specification nor the derivation of any intermediate models. In fact, it can be used even when no specification nor models are available. Since it is based on generic metrics of boundariness it can be applied even for software with non-numeric, structured, and complex inputs and\slash or outputs.  We implement our AutoBVA method as a tool and evaluate it on four different software under test (SUT).

The main contributions of this paper are:
\begin{itemize}
    \item A generic method and implementation of automated boundary value analysis (AutoBVA), and
    \item The proposal and use of a simple and very fast variant of the program derivative~\cite{feldtdobslaw2019} for efficient search, and
    \item The comparison of simple but broadly sampling and two heuristic local-search algorithm within our tool, and
    \item An empirical evaluation of AutoBVA on four SUTs to its capabilities in identifying boundary candidates.
\end{itemize}

Our results show that the proposed method can be effective even when using a simple and fast program derivative together with simple algorithms that just ensure a broad sampling of the input space. Furthermore, for some of the investigated programs, more sophisticated meta-heuristic search algorithm can provide additional and complementary value.

The rest of this paper is organized as follows. After providing a more detailed background and overview of related work in Section~\ref{sec:relwork}, we present AutoBVA in Section~\ref{AutoBVA}. The empirical evaluation is detailed in Section~\ref{sec:method} followed by the results in Section~\ref{sec:results}. The results are discussed in Section~\ref{sec:discussion} and the paper concludes in Section~\ref{sec:conclusions}. Appendix \ref{strategy_screening} and \ref{sec:apx_clustering_screening} contain details of two screening studies that supported AutoBVA meta-parameter choices for its detection and summarization phases, respectively.

\section{Related Work}
\label{sec:relwork}

In the following, we provide a brief background to boundary value analysis and the related partition testing concepts of domain testing and equivalence partitioning.

White and Cohen~\cite{white1980domain} proposed a domain testing strategy that focuses on identifying boundaries between different (sub-)domains of the input space and to ensure that boundary conditions are satisfied. As summarized by Clarke et al~\cite{clarke1982close}: ``Domain testing exploits the often observed fact that points near the boundary of a domain are most sensitive to domain errors. The method proposes the selection of test data on and slightly off the domain boundary of each path to be tested.''. This is clearly connected to the testing method typically called boundary value analysis (BVA), first described more informally by Myers in 1979 but later also included in standards for software testing~\cite{myers1979,reid1997empirical,reid2000bs,BCS2001}. Jeng and Weyuker~\cite{jeng1994simplified} even describe domain testing as a sophisticated version of boundary value testing.

While White and Cohen~\cite{white1980domain} explicitly said their goal was to ``replace the intuitive principles behind current testing procedures by a methodology based on a formal treatment of the program testing problem'' this has not lead to automated tools and a BVA is typically performed manually by human testers. Worth noting is also that while boundary value analysis is typically described as a black-box method~\cite{myers1979,reid1997empirical,BCS2001}, requiring a specification, the White and Cohen papers are less clear on this and their domain testing strategy could also be applied based on the control flow conditions of an actual implementation.

The original domain testing paper~\cite{white1980domain} made a number of simplifying assumptions such as the boundary being linear, being defined by ``simple predicates'', and that test inputs are continuous rather than discrete. While none of these limitations should be seen as fundamental they do leave a practitioner in a difficult position since it is not explicit what the method really entails when some or all of these assumptions are not fulfilled. Even though later formulations of BVA as a black-box method~\cite{BCS2001} avoid these assumptions, they, more fundamentally, do not give concrete guidance to testers on how to identify boundaries or the partitions they define. 

As one example, the BCS standard~\cite{BCS2001} states that ``(BVA) ...uses a model of the component that partitions the input and output values of the component into a number of ordered sets with identifiable boundaries.'' and that ``a partition's boundaries are normally defined by the values of the boundaries between partitions, however where partitions are disjoint the minimum and maximum values in the range which makes up the partition are used'' but do not give guidance on where to find or how to create such a partition model\footnote{An annex to the standard does provide an example of how to find partitions and identify boundaries but the specification used in the example explicitly states the boundaries so the identification task is trivial} if none is already at hand. This problem was clear already from Myers original description of BVA~\cite{myers1979} which stated ``It is difficult to present a cookbook for boundary value analysis, since it requires a degree of creativity and a certain amount of specialization to the problem at hand''. 

Later efforts to formalize BVA ideas have not addressed this. For example, Richardson and Clarke's~\cite{richardson1985partition} partition analysis method makes a clear difference between partitions derived from the specification versus from the implementation and proposes to compare them but relies on the availability of a formal specification and do not detail how partitions can be derived from it. Jeng and Weyuker~\cite{jeng1994simplified} proposed a simplified and generalized domain testing strategy with the explicit goal of automation but then only informally discuss how automation based on white-box analysis of path conditions could be done.

A very different approach is Pandita et al~\cite{pandita2010guided} that presents a white-box automated testing method to increase the Boundary Value Coverage (BVC) metric (originally presented by~\cite{kosmatov2004boundary}). The core idea is to instrument the software under test with additional control flow expressions to detect values on either side of existing control flow expressions. An existing test generation technique to achieve branch coverage, in ~\cite{pandita2010guided} the dynamic symbolic execution test generator Pex is used, can then be used to find inputs on either side of a boundary. The experimental results were encouraging in that BVC could be increased with 23\% on average and also lead to increases (11\% on average) in the fault-detection capability of the generated test inputs.

There has been several studies that empirically evaluate BVA. While an early empirical study by Hamlet and Taylor~\cite{hamlet1990partition} found that random testing was more effective, it's results were challenged in later work~\cite{reid1997empirical,yin1997automatic}. Reid~\cite{reid1997empirical} investigated three testing techniques on a real-world software project and found that BVA was more effective at finding faults both than equivalence partitioning and random testing. Yin et al \cite{yin1997automatic} compared a method based on checkpoints, manually encoding qualitatively different aspects of the input space, combined with antirandom testing~\footnote{A form of diversity-driven test generation that also relates clearly to what is recently more commonly referred to as adaptive random testing (ART).} to different forms of random testing and found the former to be more effective. The checkpoint encoding can be seen as a manually derived model of important properties of the inputs space and thus, indirectly, defines, potentially overlapping, partitions.

Even recent work on automating partition and boundary value testing has either been based on manual analysis or required a specification\slash model to be available. Hubner et al~\cite{hubner2019experimental} recently proposed a novel equivalence class partitioning method based on formal models expressed in SysML. An SMT solver is used to sample test inputs either inside or on the border of identified equivalence classes. They compared multiple variants of their proposed technique with conventional random testing. The one that sampled 50\% of test inputs within and 50\% on the boundaries between the equivalence partitions performed best as measured by mutation score. 

Related work on modeling the input space has also been done as a means to improve combinatorial testing. Boraznajy et al~\cite{borazjany2013input} proposed to divide the problem into two phases where the first models the input structure while the latter models the input parameters. They propose that ideas from partition testing can be used for the latter stage. However, for analysing the input structure they propose a manual process that can support only two types of input structures, either flat (e.g. for command line parameters that have no obvious relation) or graph-structured (e.g. for XML files for which the tree structure can be exploited).

We have previously proposed a family of metrics to quantify the boundariness of pairs of software inputs~\cite{feldtdobslaw2019}. This is a generalization of the classical definition of functional derivatives in mathematics, that we call program derivatives. Instead of using a standard subtraction (``-'') operator to measure the distance between inputs and outputs we leverage general, information theoretical results on how to quantify distance and diversity. We have previously used such measures for increasing and evaluating the benefits of test diversity~\cite{feldt2008searching,feldt2016tsd}. %RF: Add BVE study here or below.
In a recent study, we used the program derivatives to explore input spaces and visualise boundaries for testers and developers~\cite{dobslaw2020boundary}. Here, we automate this approach by coupling it to search and optimization algorithms.

In summary, early results on partition and boundary value analysis\slash testing require a specification and do not provide detailed advice or any automated method to find boundary values. One automated method has been proposed but instead requires a model of the system under test. One other method can automatically increase a coverage metric for boundary values but is white-box and requires both instrumentation of the software under test as well as advanced test generation based on symbolic execution. In contrast to existing research we propose an automated, black-box method to identify boundary candidates that is simple and straightforward to implement. 

\section{Automated Boundary Value Analysis}
\label{AutoBVA}

\begin{figure}
    \centering
    \includegraphics[width=\textwidth]{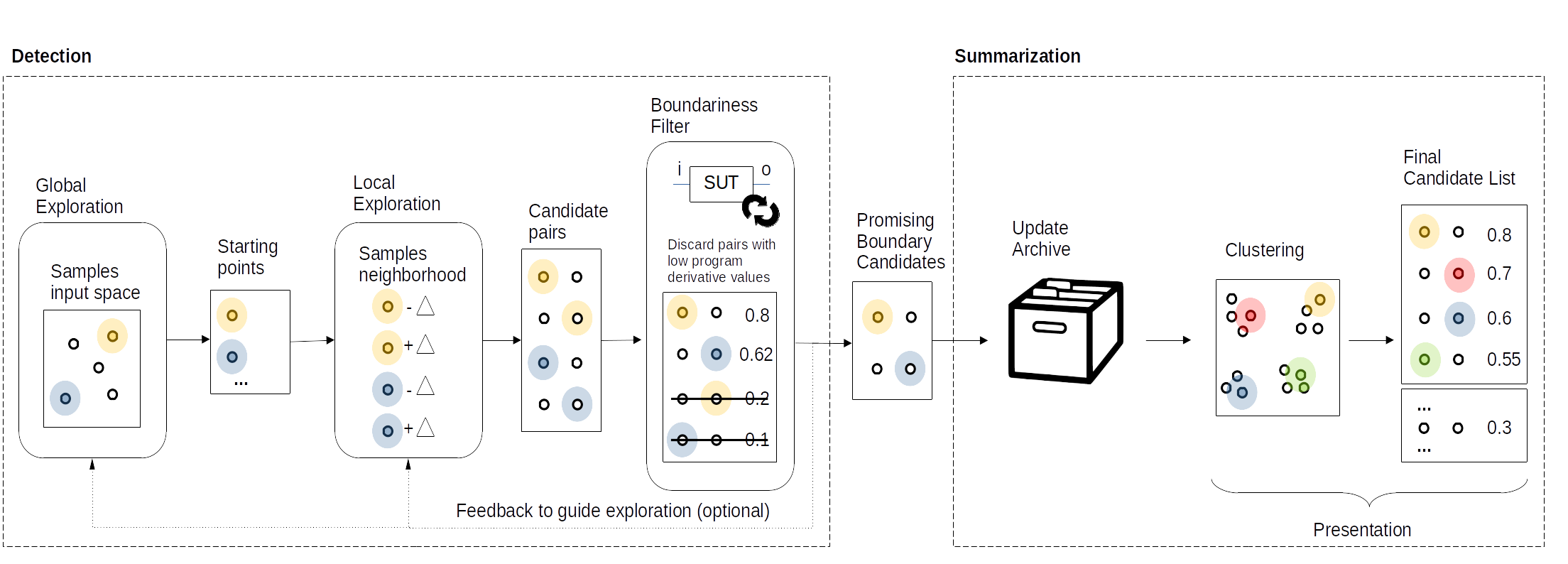}
    \caption{AutoBVA framework for automated boundary value analysis.}
    \label{fig:AutoBVE_overview}
\end{figure}

We propose to automate boundary value analysis by a fully automated detection method which outputs a set of input\slash output pairs that are then summarized and presented to testers. An overview of our proposed approach can be seen in Figure~\ref{fig:AutoBVE_overview}. The two main parts are \textit{detection} (on the left), which produces a list of promising boundary candidates, which are then \textit{summarized} and presented to the tester (on the right). The boundary value \textit{detection} method is based on two key elements: (1) \textit{search to explore} (both globally and locally) the input domain coupled with (2) the \textit{program derivative} to quantify the boundariness of input pairs. While exploration acts as a generator of new boundary candidate input pairs the program derivative acts as a filter and selects only promising candidate pairs for further processing. For summarization, an archive is updated with the new pairs and only the unique and \emph{good} boundary candidates are kept. The final list of promising candidates, in the archive, is then summarized and presented to the tester who can select the most interesting and meaningful ones and turn them into test cases. For the summarization step we here propose to use clustering to avoid showing multiple candidates that are very similar to each other.

Below we describe the three main parts of our approach: selection, search\slash exploration, and summarization. We start with selection, since that is the critical step, and use a simple SUT, \verb|bytecount|, to exemplify each step.

\subsection{Selection: Program Derivative}

We argue that the key problem in boundary value analysis is how to judge what is a boundary and what is not. If we could quantify how close to a boundary a given input is we could then use a multitude of different methods to find many such inputs. Together such a set could be used to indicate where the boundaries of the software are.

In previous work~\cite{feldtdobslaw2019}, we proposed the \textit{program derivative} for doing this type of boundariness quantification. The idea is based on a generalization of the classic definition of the derivative of a function in (mathematical) analysis. In analysis, the definition is typically expressed in terms of one point, $x$, and a delta value, $h$, which together define a second point after summation. The derivative is then the limit as the delta value approaches zero:

	$$\lim_{h\to 0} \frac{f(x+h) - f(x)}{(x+h) - x} = \lim_{h\to 0} \frac{f(x+h) - f(x)}{h}$$

We can see that this measures the \textit{sensitivity to change} of the function given a change in the input value. A large (absolute value of a) derivative indicates that the function changes a lot even for a very small change of the input. If the function here instead was the software under test, and the small change in inputs would cross a boundary it is reasonable that the output would also change more than if the change did not cross a boundary. We could then use this \textit{program derivative} to screen for input pairs that are good candidates to cross boundaries of a program.

Key to generalize from mathematical functions to programs is to realize that programs typically have many more than one input and that their types can be very different from numbers. Also, there can be many outputs and their types might differ from the types of the inputs. Instead of simply using subtraction (``$-$'') both in the numerator and denominator we need two distance functions, one for the outputs ($d_o$) and one for the inputs ($d_i$). Also, rather than finding the closest input, to actually calculate the derivative of a single input, for our purposes here we only need to quantify the \textit{boundariness} of any two, individual inputs. We thus define the \textit{program difference quotient} (PDQ) for program $P$ and inputs\footnote{We here assume each input corresponds to a single argument of the program but by using distance functions that can handle also multiple arguments the formulation becomes fully general.} $a$ and $b$ as~\cite{feldtdobslaw2019}:

\begin{equation*}
PDQ_{d_o,d_i}(a, b) = \frac{d_o(P(a), P(b))}{d_i(a, b)}
\end{equation*}

where $P(x)$ denotes the output of the program for input $x$. 

Since the PDQ is parameterized on the input and output distance functions this defines not a single but a whole family of different measures. A tester can choose distance functions so as to capture meaningful differences in outputs and\slash or inputs. In the original program derivative paper~\cite{feldtdobslaw2019}, the authors argued for compression-based distance functions as a good and general choice. However, a downside with these are that they can be relatively costly to calculate which could be a hindrance when used in the context of a search-based loop which will potentially require a very large number of distance calculations. Also, compression-based distances using mainstream string compressors such as \verb|zlib| or \verb|bzip2| might not work well for short strings, as commonly seen in testing. 

In this work, we thus focus on using one of the least costly output distance functions one can think of: \verb|strlendist| the difference in length of outputs when printed as strings. Obviously, this distance function works regardless of the type of the outputs involved. A downside is that it is coarse-grained and will not detect smaller differences in outputs of the same length. Still, if a simple and fast distance function can suffice to identify boundaries this can be a good baseline for further investigation. Also, our framework is flexible and can also use multiple distance functions for different purposes in its different components.
For example, one could use \verb|strlendist| during search and exploration while using also more fine-grained NCD-based output distance when updating the archive or during summarization (see Figure \ref{fig:AutoBVE_overview}).

For the input distance function this will typically vary depending on the software under test. If inputs can be represented as numbers or a vector of numbers we can use simple substraction or a vector distance functions such as Euclidean distance. For more complex input types one can use string-based distance functions like Normalised Compression Distance (NCD)~\cite{feldt2008searching,feldt2016tsd,feldtdobslaw2019} or even simpler ones like Jaccard distance~\cite{jaccard1912distribution}.

\subsubsection{Example: program derivative for bytecount}

We will exemplify our framework with the simple \verb|bytecount| function that is one of the most copied snippets of code on Stack Overflow but also known to contain a bug~\cite{Baltes2019,Lundblad2019}. It translates an input number of bytes into a human-readable string with the appropriate unit, e.g. ``MB'' for megabytes etc. For example, for an input of $2099$ it returns the output ''2.1 kB`` and for the input $9950001$ it returns ''10.0 MB``.

Table~\ref{tab:bytecount_someexamples} shows a set of manually selected examples of boundary candidate pairs, using a single input distance function (subtraction) but two different output distances: \verb|strlendist| and \verb|Jaccard(1)|, the Jaccard distance based on 1-grams~\cite{jaccard1912distribution}. The Jaccard distance can be used as an approximation of compression distances but is fast to calculate and applicable also for short strings. Correspondingly, in our example table, there are two different PDQ values and we have sorted in descending order based on the $PDQ_2$, i.e. that uses the $Jaccard(1)$ output distance function.

Starting from the bottom of the table, on row 6, the \verb|bytecount| output is the same for the input pair $(99948, 99949)$. This leads to PDQ values of $0.0$ regardless of the output distance used. The PDQ values are zero also for the example on row 5; even though the output has changed compared to row 6 they are still the same within the pair. We are thus in a different partition, since the outputs differ from the ones on row 6, but we are not crossing any clear boundary and the PDQ values are still zero.

The example on row 4 does show a potential boundary crossing. Even though the input distance is now higher, at 2, the outputs differ so both PDQ values are non-zero. This example pair can be improved further though, by subtracting $1$ from $99951$ to get the pair $(99949, 99950)$ shown on row 3. Since the denominator in the PDQ calculation is smaller, the PDQ value is higher and we consider it to be a better boundary candidate. In fact this is the input pair with the highest PDQ value of the ones that has these two outputs; it can thus be considered the optimal input pair to show the ``99.9 kB'' to ``100.0 kB'' boundary.

Finally, the examples on rows 2 and 1 show input pairs for which the two PDQ measures used differ in their ranking. While we can agree that both of these input pairs identify boundaries it is possible that different testers might have different preferences on which one is the more preferred. Given that $PDQ_1$, using the \verb|strlendist| output distance, is so simple and quick to compute we will use that in the experiments of this study. Future work can explore trade-offs in the choice of distance functions as well as how to combine them. Note that regardless of how the input pairs have been found they can be sorted and selected using different output distance functions when presented.

\begin{table}
\footnotesize
    \centering
\caption{Example of six boundary candidates for the bytecount SUT with their corresponding Program Difference Quotient (PDQ) values for two different output distances.}
\label{tab:bytecount_someexamples}
\begin{tabular}{rrrrrrrrrr}
\toprule
\textbf{Row} & \textbf{Input 1} & \textbf{Input 2} & \textbf{Output 1} & \textbf{Output 2} & \textbf{$d_{o1}$ (strlendist)} & 
\textbf{$d_{o2}$ (Jaccard(1))} &
\textbf{$d_{i}$} &
\textbf{$PDQ_{1}$} &
\textbf{$PDQ_{2}$}\\
\midrule
1 & 9 & 10 & 
9B & 10B &
1 & 0.75 & 1 & 1 & 0.75\\

2 & 999949999 & 999950000 & 
999.9 MB & 1.0 GB & 
2 & 0.63 & 1 & 2 & 0.63\\

3 & 99949 & 99950 & 
99.9 kB & 100.0 kB & 
1 & 0.43 & 1 & 1 & 0.43 \\

4 & 99949 & 99951 & 
99.9 kB & 100.0 kB & 
1 & 0.43 & 2 & 0.50 & 0.21 \\

5 & 99951 & 99952 & 
100.0 kB & 100.0 kB & 
0 & 0.0 & 1 & 0.0 & 0.0 \\

6 & 99948 & 99949 & 
99.9 kB & 99.9 kB & 
0 & 0.0 & 1 & 0.0 & 0.0 \\
\bottomrule
\end{tabular}
\end{table}

\subsection{Exploration: Generation of Candidate Pairs}
\label{exploration}
While the program derivative can help evaluate whether an input pair is a good candidate to detect a boundary, there is a very large number of possible such pairs to consider. Thus, we need ways to explore the input space and propose good candidate pairs, i.e. that have high program derivative values. A natural way to approach this is as a search-based software engineering problem~\cite{harman2001search,afzal2009systematic,feldt1998generating}, with the program derivative as the goal\slash fitness function and a search algorithm that tries to optimize for higher program derivative values.

However, to identify the boundaries it is not enough to simply find and return one candidate pair. Most software will have multiple and different boundaries in their input domain. Furthermore, boundaries are typically stretched out over (consecutive) sets of inputs. 
%So having only a single, best candidate pair would not be sufficient. 
The search and exploration procedure we chose should thus output a set of input pairs that are, ideally, spread out over the input domain (to find multiple boundaries) as well as over each boundary (to help testers identify where it is).

An additional concern when using search-based approaches is the shape of the \textit{fitness landscape}, i.e. how the fitness value changes over the space being searched~\cite{smith2002fitness,yu2006through}. Many search and optimisation approaches assume or at least benefit from a smooth landscape, i.e. where small steps in the space lead to only a small change in the fitness value. It is not clear that we can assume this to be the case for our problem. The program derivative might be very high right at the boundary while showing very little variation when inside the partitions on either side of the boundary.
Worst case, this could be a kind of needle-in-a-haystack fitness landscape \cite{yu2006through} where there is little to no indication within a sub-domain to guide a search towards its edges, where the boundary is.

Given these differences compared to how search and optimisation has typically been applied in software testing we formulate our approach in a generic sense. We can then instantiate it using different search procedures and empirically study which ones are more or less effective. However, given that we are searching for a pair of two inputs we separate the search component into two steps: a global and a local exploration\slash search step. 

\begin{algorithm}
        \caption{Automated Boundary Detection - AutoBD-2step}
        \begin{algorithmic}[1]
        \label{alg:AutoBD}
            \REQUIRE Software under test $SUT$, Boundariness quantifier $Q$, Sampling strategy $SS$, Boundary search $BS$
            \ENSURE Boundary candidates $BC$
            \STATE $BC = \emptyset$
            \WHILE{stop criterion not reached}
                \STATE $input = SS.sample(SUT, BC)$ %rf: I added BC as parameter here since that makes the global search potentially much more powerful, even if we don't explore that here.
                \COMMENT{globally sample a starting point}
                \STATE $PBC = BS.search(SUT, Q, input)$ \COMMENT{locally explore and detect potential candidate(s)}
                \STATE $BC = BC \cup \{c | c \in PBC \wedge Q.evaluate(c) > threshold(BC) \}$
            \ENDWHILE
            \RETURN $BC$
        \end{algorithmic}
\end{algorithm}

Algorithm \ref{alg:AutoBD} outlines this generic, 2-step method for automated boundary detection. Given a software under test ($SUT$), it returns a set of boundary candidate pairs ($BC$). It has three additional parameters: a way to quantify the boundariness of pairs of inputs ($Q$), a (global) sampling strategy ($SS$) to propose starting points for local exploration, and a (local) boundary search ($BS$) strategy. These exploration strategies capture two different types of sampling\slash search procedures. The global one explores the input space of the SUT as a whole by sampling different input points (line 3). The local strategy will then search from the starting point (line 4) and try to identify new potential boundary candidate pairs according to the boundariness quantifier.

Each of the potential new candidates (in $PBC$) are then evaluated and their boundariness compared to a threshold (line 5) and added to the final set ($BC$) returned. The \verb|threshold| function simply captures the fact that we might not use a fixed threshold but rather can allow more complex updating schemes where the threshold is based on the candidates that have already been found. 

We have split the search and exploration procedure into two for clarity and since they have two slightly different goals. The goal of the global sampling strategy is to propose inputs that have not been previously considered (to avoid wasting effort) and, ideally, that are likely to be close to a boundary. The goal for the local search strategy is rather to ensure that only inputs that have a small distance to the starting input are considered. This division of labor can thus increase the likelihood that we consider only input pairs that are close to each other, i.e. neighbors. Since this input distance value is in the denominator of the program derivative the use of the local search strategy thus helps ensure the denominator is small, which should lead to higher derivative values. 

Since the scale of the derivative values that the algorithm is likely to encounter cannot always be known (it will depend on the specific distance functions used) it is natural to decide the threshold (Algorithm \ref{alg:AutoBD}, line 5) dynamically. For example, the threshold could be taken as some percentile (say, 90\%) of the boundariness values of the candidate set saved so far. Alternatively, even more elaborate boundariness testing and candidate set update procedures can be used. For example, rather than simply adding to the current set whenever a sufficiently good boundary value is found, we could save a top-list of the highest boundariness values found so far. The update would then be generalized so that it can also delete previously added candidate pairs that are no longer promising.

As a simple example, consider a $SUT$ that takes a single, integer number as input. One variant of our general framework could be to choose random sampling of integers as the global exploration strategy and use the addition and subtraction of the values one ($1$) or two ($2$) in the local search step. If the global strategy sampled the integer $10$, the boundariness of four different candidate pairs would be probed in the local search: $(10, 11)$, $(10, 9)$, $(10, 12)$, and $(10, 8)$. If the $SUT$ in question was actually the above mentioned \verb|bytecount| PBC would then return the candidate pair $(10, 9)$ since it has the highest PDQ value.

While the benefits of creating candidate input pairs in two steps may not be obvious for this particular example it is easy to imagine the local search step being more important when the inputs are of a complex and structured datatype, such as for example an XML tree. If the search was done with a single, global exploration step it would be relatively unlikely that the two inputs of a candidate pair would have a small distance. By doing this in two steps, we could first sample one XML tree and then explore its neighborhood in the input space by small mutations to it. However, for future work, we don't rule out exploring other ways to structure the search\slash exploration step; a single step that searches directly for a pair with high program derivative is important future work.

In the experiments of this paper we will use one and the same (global) sampling strategy to sample starting points but compare two different boundary search algorithms, namely Local Neighbor Sampling and Boundary Crossing Search. Below we describe these different components in more detail.

\subsubsection{Global Sampling}
\label{sec:global_sampling}

All inputs in the SUTs investigated in this paper are numbers. While we could use uniform sampling in valid ranges to globally sample (Algorithm \ref{alg:AutoBD}, line 3) such numbers we argue that in general it is better to sample uniformly over the \textbf{bits}, i.e. \textit{bituniform} sampling, making up the numbers; since there are many more large numbers, standard uniform sampling would otherwise tend to favor larger numbers.

For a broad exploratory sampling we introduce a complementing technique that we call Compatible Type Sampling (CTS), i.e. the argument-wise sampling based on compatible types per argument. An example of compatible types are all integers types with specific bit size. For instance, in the Julia programming language we use in our experiments, booleans (Bool), 8-bit (Int8) and 32-bit integers (Int32) types are compatible because they all are sub-types of Integer.

More details and justification for the global sampling strategy we use here, bituniform sampling combined with CTS, can be found in Appendix \ref{strategy_screening}. We do note that in general, more advanced test input generation strategies~\cite{feldt2013} can be used, and adapting them to the SUT and its arguments will likely be important when further generalizing our framework.

\subsubsection{Local Neighbor Sampling (LNS)}

The simpler of the implementations of the here introduced local search strategies for Algorithm \ref{alg:AutoBD} (line 4), Local Neighborhood Sampling (LNS), is presented as Algorithm \ref{alg:lns}. The basic idea with LNS is to structurally sample neighboring inputs, i.e. inputs close to a given starting point $i$, to form potential candidate pairs including $i$. For that, the algorithm processes mutations over all individual arguments (line 3) considering all provided mutation operators $mos$ (line 4). For integer inputs, the mutation operators are basic subtraction and addition. Outputs are produces for the starting point (line 2) and each neighbor (line 6), to form the candidate pairs (line 7). Without filtering, they are all added to the set of potential boundary candidates (line 8), returned by the algorithm (line 11). LNS is used as a trivial baseline implementation to better understand what is possible using AutoBD without sophisticated search.

\begin{algorithm}
        \caption{\emph{search} -- Local Neighbor Sampling (LNS)}
        \begin{algorithmic}[1]
        \label{alg:lns}
            \REQUIRE Software under test $SUT$, mutation operators $mos$, Starting Point $i$
            \ENSURE potential boundary candidates $PBC$
            \STATE $PBC = \emptyset$
            \STATE $o = SUT.execute(i)$
            \FOR{$a \in arguments(SUT)$}
                \FOR{$mo \in mos[a]$}
                    \STATE $n = mo.apply(i, a)$
                    \STATE $o_n = SUT.execute(n)$
                    \STATE $c = \langle i, o, n, o_n \rangle$
                    \STATE $PBC = PBC \cup \{c\}$
                \ENDFOR
            \ENDFOR
            \RETURN $PBC$
        \end{algorithmic}
\end{algorithm}

\subsubsection{Boundary Crossing Search (BCS)}

\begin{algorithm}
        \caption{\emph{search} -- Boundary Crossing Search (BCS)}
        \begin{algorithmic}[1]
        \label{alg:bcs}
            \REQUIRE Software under test $SUT$, Boundariness quantifier $Q$, mutation operators $mos$, Starting Point $i$
            \ENSURE potential boundary candidates $PBC$
            
            \STATE $a = rand(arguments(SUT))$ \COMMENT{select random argument}
            \STATE $mo = rand(mos[a])$ \COMMENT{select random mutation operator for argument}
            \STATE $i_{next} = mo.apply(i, a)$ \COMMENT{mutate input a first time in single dimension}
            \STATE $o = SUT.execute(i), o_{next} = SUT.execute(i_{next})$ \COMMENT{produce outputs}
            \STATE $c_{init} = \langle i,o, i_{next}, o_{next} \rangle$ \COMMENT{ instantiate initial candidate}
            \STATE $\Delta_{init} = Q.evaluate(c_{init})$ \COMMENT{ calculate candidate distance}
            \STATE $c = \langle i_1, o_1, i_2, o_2 \rangle$, with $i_1$ obtained by a finite number of chained mutations $mo$ of argument $a$ over $i$, and
            \STATE \qquad \qquad \qquad \qquad \qquad$i_2 = mo.apply(i_1, a)$, and
            \STATE \qquad \qquad \qquad \qquad \qquad$o_1 = SUT.execute(i_1)$, and
            \STATE \qquad \qquad \qquad \qquad \qquad$o_2 = SUT.execute(i_2)$, and
            \STATE \qquad \qquad \qquad \qquad \qquad $\Delta_c = Q.evaluate(c)$, and
            \STATE \qquad \qquad \qquad \qquad \qquad $\Delta_c > \Delta_{init}$ or $c_{init} = c$
            \RETURN $\{c\}$
        \end{algorithmic}
\end{algorithm}

Boundary Crossing Search (BCS) is the second local search method we introduce for comparison (see Algorithm \ref{alg:bcs}). It is a heuristic algorithm that uses a boundariness quantifier $Q$, in our experiments the program derivative, to seek a single potential boundary candidate that locally stands out in comparison to starting point $i$. For a random direction (argument $a$ in line 1 and mutation operator $mo$ in line 2) a neighboring input $i_{next}$ gets mutated (line 3) and outputs for both inputs produced (line 4) to define the initial candidate (line 5) for which the boundariness gets calculated (line 6).

Lines 7-12 describe the constraints and conditions that must hold true for a resulting boundary candidate $c$ to locally stand out. This search can be implemented in a variety of ways. We implemented a binary search that first identifies the \emph{existence} of a boundary crossing by taking larger and larger steps and calculating the difference $\Delta_{c}$ to find an input for the state in which the boundariness is greater than $\Delta_{c}$, and thereby guaranteeing the necessary condition in line 12\footnote{A stop criterion returns the original point in case no difference could be picked up.}. Once that is achieved, the algorithm \emph{squeezes} that boundary to obtain $c$ which is the nearest point to $c_{init}$ that ensures there is a notable local difference in neighboring inputs, and by that guaranteeing the neighboring constraint in line 8.

\subsection{Summarization: Validity-Value Similarity Clustering}
\label{sec:summarization}

The boundary candidate set resulting from AutoBD-2step (Algorithm \ref{alg:AutoBD}) can be extensive. However, humans information processing is limited and more fundamentally many of the boundary pairs found can be similar to each other and represent the same or a very similar boundary. The summary method(s) we choose should thus not only limit the number of candidates presented, those candidates also need to be different and represent different groups of behaviors and boundaries over the input space. 

Furthermore, the goals for the summary step will likely differ depending on what the boundary candidates are to be used for; comparing a specification to actual boundaries in an implementation is a different use case than adding boundary test cases to increase coverage. Thus we cannot provide one general solution for summarization and future work will have to inspect different methods to cluster, select, prioritize but also visualise the boundary candidates. 

In the following, we propose one particular, but very general, summarization method. Our hope is that it can show several building blocks needed when creating summarization methods for specific use cases. But also that it can act as a general, fallback method that can provide value regardless of the use case. This is also the method we use in the experimental evaluation of the paper. We consider a general, black-box situation with no specification available. Thus, we only want to use information about the boundary candidates themselves. The general idea is to identify meaningful clusters of similar candidates and then sample only one or a few representative candidates per cluster and present to the tester.

For instance, Table~\ref{tab:bytecount_summary_example} contains a subset of (nine) candidate pairs found by our method for the \verb|bytecount| example introduced above. Different features can be considered when grouping and prioritizing. We see that candidates differ at least in terms of the type of their output, whether the outputs are valid return values or exceptions, as well as in the actual values of inputs and\slash or outputs themselves. For example, both outputs for the candidate on row 2 are strings and are considered normal, valid return values. On the other hand, for row 9 both outputs are (Julia) exceptions indicating that a string of length 6 ($"kMGTPE"$) has been accessed at (illegal) positions 9 and 10, respectively.

We argue that the highest level boundary in boundary value analysis is that between valid and invalid output values. Any time there is an exception thrown when the $SUT$ is executed we consider the output to be invalid, if not the output is valid\footnote{Note that in the implementation we have to clearly differentiate between the situation when an exception was thrown during execution from the one where the function itself returns a value that is an exception. Otherwise, our framework could not be used for functions that manipulate exceptions (without raising any exceptions while doing so).}. Since we consider pairs of inputs there are two outputs per candidate that can be grouped in three, what we call, \textit{validity groups}:

\begin{itemize}
    \item two valid outputs in the boundary pair (VV),
    \item one valid output and one invalid (error) output (VE), and
    \item two invalid (error) outputs (EE).
\end{itemize}

We use validity as the top-level variation dimension and produce individual summaries for each of these three groups. In Table~\ref{tab:bytecount_summary_example}, the validity group is indicated in the rightmost column (VV for lines 1-7, VE in line 8, and EE in line 9). Within each validity group we can consider other characteristics to further categorize candidates. For example, we could use the type of the two outputs to create further (sub-)groups. This is logical since it is not clear that it is always meaningful to compare the similarity of values of different types. However, for many SUTs and their validity groups the types might not provide further differentiation, since they are often the same.

In the final step we propose to cluster based on the similarity of features within the identified sub-groups. A type-specific similarity function can be used or we can simply use a general, string-based distance function after converting the values to strings. After calculating the pair-wise similarities (distances) we can create a distance matrix per sub-group. This can then be used with any clustering method (in the experiments in this paper we used k-means clustering~\cite{likas2003global}). From each cluster then we select one representative (or short) boundary pair to present to the tester. The distance matrix can also be used with dimensionality reduction methods to visualise the validity-type sub-group. We thus call our summarization method \textit{Validity-Value Similarity Clustering}.

\begin{table}
\tiny
    \centering
    \caption{A summary of boundary candidates found for bytecount by our method. Rows 1-7 are for the valid-valid (VV) validity group of type String-String, row 8 for the valid-error (VE) group of type String\slash BoundsErrror, and row 9 for error-error (EE) group of type BoundsError\slash BoundsError. The candidates of rows 1-7 are the shortest candidate pairs of each identified cluster.}
    \label{tab:bytecount_summary_example}

\begin{tabular}{rrrrrr}
\toprule
\textbf{Row} & \textbf{Input 1} & \textbf{Input 2} & 
\textbf{Out 1} & \textbf{Out 2} & \textbf{Validity}\\

\midrule
1 & false & true & falseB & trueB & VV \\

2 & 9 & 9B & 10 & 10B & VV\\

3 & -10 & -9 & -10B & -9B & VV\\

4 & 999949 & 999950 & 999.9 kB & 1.0 MB & VV\\

5 & 99949999999999999 & 99950000000000000 & 99.9 PB & 100.0 PB & VV\\

6 & 9950000000000001999 & 9950000000000002000 & 9.9 EB & 10.0 EB & VV\\

7 & -1000000000000000000000000000000 & -999999999999999999999999999999 & -1000000000000000000000000000000B & -999999999999999999999999999999B & VV\\

\midrule
8 & 999999999999994822656 &
999999999999994822657 & 1000.0 EB & BoundsError("kMGTPE", 7) & VE\\

\midrule
9 & 999999999999990520104160854016 & 999999999999990520104160854017 & BoundsError("kMGTPE", 9) & BoundsError("kMGTPE", 10) & EE\\
\bottomrule
\end{tabular}
\end{table}

\section{Empirical Evaluation}
\label{sec:method}

For evaluation, we run our framework using two different BVE search strategies (independent variable) on four different types of SUTs (control variables) and study the sets of boundary candidates returned, in detail. Our SUTs are program functions. Our focus is on evaluating to what degree the framework can find many, diverse, and high-quality candidate boundary pairs. Specifically, we address the following research questions:

\begin{itemize}
    \item RQ1: Can large quantities of boundary candidates be identified? How do the different exploration strategies compare against one another in terms of identifying unique boundary candidates?
    \item RQ2: Can the method robustly identify boundary candidates that cover a diverse range of behaviors? 
    \item RQ3: To what extent can AutoBVA reveal relevant behavior or potential faults?
\end{itemize}

Through RQ1 we try to understand to what extent AutoBVA can pick up \emph{potential} boundary candidates (PBC), comparing two local search strategies. We compare both the overall quantities and quantities of \textit{uniquely} identified PBCs using a basic boundary quantifier. Uniqueness here is measured in relation to the set of all PBCs for a SUT over all experiments irrespective of the local search applied.

Through RQ2, we try to understand how well AutoBVA covers the space of possible boundaries between equivalence partitions in the input space of \emph{varying} behavior. For a given arbitrary SUT, we argue that there is no one-size-fits-all approach to extracting\slash generating ``correct'' equivalence partitions; many partitions and, thus, boundaries exist and which ones a tester consider might depend on the particular specification, the details with which it has been written, the interests and experience of the tester etc. Therefore, we use our summarization method (Validity-Value Similarity Clustering as described in Section~\ref{sec:summarization}) to group similar PBCs within each validity group (VV, VE and EE). We answer RQ2 by comparing how the PBCs found by each exploration method cover those different clusters. Comparing the coverage of these clusters allows us to interpret the behaviour triggered by the set of PBCs. For instance, even though two boundary candidates identified for the Date constructor SUT, bc1 = (\textbf{28}-02-2021, \textbf{29}-02-2021) and bc2 = (\textbf{28}-02-2022, \textbf{29}-02-2022), are different they do cover the same valid-error boundary\footnote{Note that both pairs belong to the valid-error (VE) validity group as both 2021 and 2022 are \textbf{not} leap years and thus February 29th leads to an ArgumentError exception being thrown.}. It is thus not clear that finding them both, or many similar boundary candidates showing a similar boundary, helps in identifying diverse boundary behavior.

The more quantitative approaches we use for RQ1 and RQ2 cannot really probe whether the candidates found are of high quality, i.e. if the boundaries they indicate are ones that are important to test. For RQ3, we thus perform a qualitative analysis of the identified PBCs by manually investigating each cluster, systematically sampling candidates with varied output length in each and analysing them. Consequently, we analyse whether and how often AutoBVA can identify relevant, rare and\slash or fault-revealing SUT behaviour. 

In the following, we detail how we setup each stage of AutoBVA in our experiments.

\subsection{Setup of Selection and Exploration Step}

We perform an experiment to gather data to then answer our research questions. We analyse AutoBVA by contrasting two exploration strategies (one factor - two levels) applied to four chosen SUTs (control variables). Prior to the main experiment we performed two screening studies to configure (1) the (global) sampling strategy and (2) the clustering of boundary candidates (see Appendices \ref{strategy_screening} and \ref{sec:apx_clustering_screening}).

The sampling method is fixed as the one best performing configuration from the screening study and  uses both bituniform sampling and CTS for all experiments (see Appendix \ref{strategy_screening}). For each SUT, we conducted two series of runs, one short for 30 seconds each and one longer for 600 seconds (10 minutes), to understand the convergence properties of AutoBVA. While these running times are ad hoc, we selected them to at least loosely correspond to more direct (30 seconds) and more offline (10 minutes) use by a tester. The boundariness quantifier for all experiments is \emph{strlengdist}, as output distance, i.e. the length of the output when seen as a string. Since all input parameters of the SUTs in this study are integer types, numerical distance is used (implicitly) as input distance, and we use mutation operators increment (add to integer) and decrement (subtract from integer) during search and exploration. We repeat each experiment execution ten times to account for variations caused by the pseudorandom number generator used during search. Table \ref{tab:configs} summarises the setup of the experiment. 

\begin{table}
    \centering
    \caption{The different configurations used to run each of the four SUTs.}
    \label{tab:configs}
    \begin{tabular}{ll}
    \toprule
    \textbf{Parameter}         & \textbf{Choice of values}  \\
    \midrule
    Exploration strategy (ES)       & LNS, BCS                \\
    SUTs & \verb|bytecount|, BMI-Value, BMI-Class, Date           \\
    Sampling method (SS) & bituniform + CTS activated    \\
    Boundariness quantifier (PD) & \emph{strlendist}      \\
    Threshold & 0 \\
    Mutation operators  (m)  & increment\slash decrement (++\slash - -)\\
    Stop criterion & \{30,600\} seconds \\
    \bottomrule
    \end{tabular}
\end{table}

\subsection{Setup of Summarization Step}
\label{sec:summarization_detials}

The dependent variables in our experiment are the number of unique candidates found (RQ1), the number of clusters covered (RQ2), and the characteristics of interesting candidates (RQ3) found by each exploration approach. To summarize a large set of boundary candidates we have to decide and extract a set of complementing features able to group similar candidates and set apart those that are dissimilar. Ideally we want to have a generic procedure for this, which can give good results for many different types of SUTs.

The program derivative seems a good candidate for a feature that helps separating by candidate differences within boundaries. Another seemingly relevant aspect to be captured as a feature was the degree of similarity of a boundary candidate outputs in relation to the overall set of outputs in the entire set of candidates, i.e. the uniqueness of an output. Essentially, we want to select features so that boundary pairs with similar outputs are grouped together. The choice of features is formalized below.

We implement the AutoBVA summarization by Validity-Value Similarity Clustering using k-means clustering~\cite{hartigan1979algorithm}. We choose k-means clustering because it is one of the simplest, well-studied, and understood clustering algorithms and is also widely available~\cite{xu2015comprehensive}. Clustering was done per validity group to avoid mixing pairs that have very different output types, namely: VV (String, String), VE (String, ArgumentError), and EE (ArgumentError, ArgumentError).

We span a feature-space over the boundary candidates to capture a diverse range of properties and allow for a diversified clustering from which to sample representatives per cluster. 

We extract features from the output differences between boundary candidates, since input distances within the boundary candidates are already factored into the selection of the candidates. 
Moreover, outputs can easily be compared in their ``stringified'' version using a generic information theoretical measure $Q$, typically a string distance function. 

Our goal is that the features that span the space shall be generic and capture different aspects of the boundary candidates. We therefore introduce two feature types: (1) WD captures the \textit{differences within} a boundary candidate as the distance between the first and the second output; (2) U is a two-attribute feature that captures the \textit{uniqueness} of a candidate based on the distance between the first ($U_1$) and second ($U_2$) output to the corresponding outputs of all other candidates in the set.\footnote{All distance values from $Q$ are normalised between zero and one keep all features and corresponding attributes on the same scale.} Considering $Q$ as the distance measure chosen for the outputs, we define U and WD, for a boundary candidate $j \in BC, j = (out_{j1}, out_{j2})$ as:

\begin{equation*}
\begin{split}
WD_{j} & = Q(out_{1j}, out_{2j}) \\
\\
U_j  & = (U_{j1},U_{j2});\mathrm{\ where\ } 
U_{jk} = \sum_{j' \in BC} Q(out_{jk}, out_{j'k}),\ \mathrm{k = 1, 2}
\end{split}
\end{equation*}

To understand which combination of distance measures ($Q$) yields better clustering of boundary candidates, we conducted a screening study using three different string distances to measure the distance between the outputs, namely, \emph{strlendist}, Levensthein (Lev) and Jaccard of length two (Jacc). These common metrics cover different aspects of string (dis)similarity, each with their own trade-off. For instance, \emph{strlendist} is efficient but ignores the characters of both strings, whereas Jacc compares combinations of characters but disregards specific sequences in which those characters show up (e.g., missing complete names or identifiers); lastly, Lev is the least efficient but more sensitive to differences between the strings. Nonetheless, all three measures only consider lexicographic similarity and are not sensitive to semantics, such as synonyms or antonyms.

For simplicity, the screening study was done only on the \verb|bytecount| SUT. Details of the screening study and examples of features extracted from boundary candidates are presented in Appendix \ref{sec:apx_clustering_screening}. Our screening study reveals that \emph{strlendist}(WD) and Jaccard(WD, U) is the best combination of features and distance measures that yield clusters of good fit with high discriminatory power. Choosing those types of models yields more clusters that can be differentiated with high accuracy, hence allowing for a more consistent comparison of cluster coverage between exploration strategies. Moreover, clearer clusters are also useful in practice as they allow AutoBVA to suggest testers with more diverse individual boundary candidates.

Formally, we create a feature Matrix $M$ over boundary candidates of a SUT with each row $i$ representing each attribute from the features over each boundary candidate $j$, as defined below.

$$ M = \begin{bmatrix}
WD\\
U
\end{bmatrix},$$

For this experiment, each $M$ has four rows, one per attribute in the chosen features: \emph{strlendist}(WD), Jaccard(WD), Jaccard($U_1$) and Jaccard($U_2$). The number of columns ($j$) vary depending on the number of candidates found per exploration approach and SUT. We run the clustering 100 times on each SUT and their corresponding feature matrices $M$. To evaluate the coverage of the boundary candidates (RQ2), we choose the clustering discriminating best, i.e. the one resulting in most clusters based on the top five percentile Silhouette scores.

We reduce the risk of cluster-size imbalance, and thereby improve on overall clustering quality, by selecting only a diverse subset of boundary candidates for the clustering and assign the remaining boundary candidates to the clusters with most similar diversity centers. For that we create an initial diversity matrix as of above using 1000 randomly selected candidates\footnote{For SUTs with fewer than that, we skip this step.}. We then remove the bottom 100 in terms of overall diversity (based on the sum of all normalized diversity readings) and redo this by substituting them with 100 of the remaining candidates. Until no more candidates exist, we repeat this step to receive M for clustering. 

\subsection{Description of SUTs}
We investigate four SUTs, namely: a function to print byte counts, Body Mass Index (BMI) calculation as value and category, and the constructor for the type Date. Those SUTs are comparable (i.e., unit-level that have integers as input) but each has peculiar properties and different sizes of input vectors. For instance, when creating Dates, the choice of months affect which are the valid range for days (and vice-versa), whereas the result of a BMI classification depends on the combination of both input values (height and weight). Below, we explain the input, output and reasoning for choosing each SUT. The code for each SUT is available in our reproduction package\footnote{\url{https://github.com/feldob/repro_autobva}}.

\verb|bytecount(i1: Int)|: Receives an integer representing a byte value (i1). The function returns a human readable string (valid) or an exception signalling whether the input is out of bound (invalid). The largest valid input are those representable as Eta-bytes. Chosen for being the most copied code in StackOverflow. Moreover, the code has a fault in the boundary values when switching between scales of bytes (e.g., from 1000 kB to 1.0 MB).

\verb|bmi_value(height: Int, weight: Int)|: Receives integer values for a person's height (in cm) and weight (in kg). The function returns a floating point value resulting from $ weight / (height / 100)^2$ (valid) or an exception message when specifying negative height or weight (invalid). Chosen because the output is a ratio between both input values, i.e., different combinations of height and weight yield the same BMI values.

\verb|bmi_classification(height: Int, weight: Int)|: Receives integer values for a person's height (in cm) and weight (in kg). Based on the result of \verb|bmi_value(height: Int, weight: Int)|, the outcome is a string indicating whether the person is Underweight, Healthy, Overweight or Obsese (valid). Otherwise, it returns an exception caused by negative height or weight (invalid). Chosen because the boundaries between classes depend on the combination of the input values chosen, leading to a variety of valid and invalid combinations.

\verb|date(year: Int, month: Int, day: Int)|: Receives an integer value representing a year, month and day. The function returns the string for the specified Date in the proleptic Gregorian calendar (valid).\footnote{We use the constructor from the Julia language as in \url{https://docs.julialang.org/en/v1/stdlib/Dates/}.} Otherwise, it returns specific exception messages for incorrect combinations of year, month, and day values (invalid). Chosen because Date has many boundary cases that are conditional to the combination of outputs (e.g., the maximum valid day value vary depending on the month, or the year during a leap year).

Our choice of SUT offers a gradual increase in the cognitive complexity of input, where the tester needs to understand (1) individual input values, (2) how they will be combined according to the specification, and (3) how changing them impacts the behaviour of the SUT. For instance, when choosing input for year for the Date constructor, a tester can simply choose arbitrary integer values (case `1') or think of year values that interact with February, 29 to check leap year dates (case `2'). Another example would be choosing test input for BMI, in which the tester needs to manually calculate specific combinations of height and weight to verify all possible BMI classifications (case `3'). In parallel to all those cases a tester would need to check the boundaries for the types used (e.g., maximum or minimum integer values). Note that systematically thinking of inputs to reveal boundaries is a multi-faceted problem that depends on the SUT specification (e.g., what behaviours should be triggered), the values acceptable for the input type and the output created independently of being a valid outcome or an exception.

\section{Results and Analysis}
\label{sec:results}

Below we present the results and analyse them in relation to each of the research questions. The next section then provides a more general discussion on lessons learnt, implications, as well as future work.

\subsection{RQ1 - Boundary candidate quantities}

Table \ref{tab:direct_stats} summarises the number of common and unique boundary candidates found by the two search strategies LNS and BCS. For each SUT and search strategy it shows result for the 30 second and the 600 second runs, individually. For each time control the mean and standard deviation, over the 10 repetitions, of the number of potential boundary candidates as well as the number of unique candidates is listed. For example, we can see that BCS in a 600 second run for the \verb|Date| SUT finds on average 897.4 +/- 82.6 PBCs out of a total of 45456 unique ones (found over all runs and search strategies). And overall, 7276 of the total 45456 PBCs were uniquely only found by BCS, i.e. in none of the LNS runs any of these 7276 were found.

We see that overall a large number of boundary candidates can be found by both methods. Except for \verb|bytecount|, there is also a large increase in the number of candidates found as the execution time increases. While the number of candidates found per second does taper off also for \verb|BMI-class|, \verb|BMI| and \verb|Julia Date| (from 166, 552, and 81 per second for the short runs to 117, 380, and 76 for the long runs, respectively), longer execution clearly has the potential to identify more candidates. Since the 20 times longer execution time, for \verb|bytecount|, only finds one additional candidate (58 total versus 57 for 30 seconds) it might be useful to terminate search and exploration when the rate of new candidates found goes below some treshold.

For the \verb|bytecount| SUT, only BCS finds a unique set of candidates, meaning that all boundaries identified by LNS were also identified by BCS. This means that only 14 (58-44) of the all candidates found was found by LNS, even after 10 runs of 600 seconds, and all of those were also found by BCS. For the other SUTs, LNS clearly finds more candidates and more unique candidates, between 5 and 15 times more, depending on the SUT.

Thus, overall, LNS produces a higher quantity of boundaries. This is expected since it is essentially a random search strategy with minimal local exploration. The effect can likely be explained by two reasons that may also interplay. First, the low algorithmic overhead of the LNS search method enables it to make more calls to the underlying SUT given a fixed time budget. Second, a proportion of the BCS searches can fail and return no boundary candidates at all since the input landscape does not regularly lead to changes in output partition by means of single-dimensional, i.e. to one input, mutations. However, the quantity of boundary candidates might not directly translate to finding more diverse and ``better'' boundary candidates; next we thus need to consider RQs 2 and 3.

\begin{table}
    \centering
    \caption{Descriptive statistics $(\mu \pm \sigma)$ over the potential boundary candidates (PBC) found by both BCS and LNS. Total refers to the size of the union set of the candidates found during the 20 executions of each strategy.}
    \label{tab:direct_stats}
    \begin{tabular}{@{}llrrrrrr@{}}
    \toprule
    \textbf{SUT} & \textbf{Strategy} & \multicolumn{3}{c}{\textbf{30 seconds}} & 
    \multicolumn{3}{c}{\textbf{600 seconds}}\\
    \cmidrule(lr){3-5} \cmidrule(lr){6-8}
       &    &  \textbf{Total} & \textbf{\# PBC found} &  \textbf{\# Unique} & \textbf{Total} & \textbf{\# PBC found} & \textbf{\# Unique} \\
\midrule
\verb|bytecount| & LNS & 57 & $9.5 \pm 0.7$ & 0 & 58 & $12.0 \pm 0.7$ & 0 \\
 & BCS & 57 & $56.1 \pm 0.6$ & 46 & 58 & $57.2 \pm 0.4$ & 44 \\
    \midrule
BMI-class & LNS & 4979 & $623.9 \pm 40.5$ & 4194 & 70265 & $8233.2 \pm 196.6$ & 59890 \\
 & BCS & 4979 & $104.2 \pm 9.3$ & 543 & 70265 & $1398.7 \pm 45.2$ & 7631 \\
\midrule
BMI & LNS & 16549 & $2054.0 \pm 61.3$ & 14918 & 227899 & $27581.0 \pm 265.4$ & 205389 \\
 & BCS & 16549 & $209.3 \pm 43.0$ & 1126 & 227899 & $2959.1 \pm 52.9$ & 15895 \\
\midrule
Julia Date & LNS & 2444 & $246.1 \pm 13.7$ & 2232 & 45456 & $4351.6 \pm 86.3$ & 37216 \\
 & BCS & 2444 & $21.6 \pm 5.9$ & 191 & 45456 & $897.4 \pm 82.6$ & 7276 \\
    \bottomrule
    \end{tabular}
\end{table}

\vspace{0.4cm}
\begin{mdframed}[style=style1]
\textit{Key findings (RQ1):}
Large quantities of boundaries can be identified by both exploration strategies. Overall, Local Neighbor Search (LNS) finds more candidates and more unique candidates than Boundary Crossing Search (BCS) for the more complex SUTs with multiple input arguments, while BCS finds larger numbers and more unique candidates for the one-input SUT.
\end{mdframed}

\subsection{RQ2 - Robust coverage of diverse behaviors}

Using our Validity-Value Similarity Clustering method, we obtained between 5 and 15 clusters across the different SUTs and validity groups (VV, VE and EE). We summarise the coverage of those clusters per SUT, exploration strategy, and execution time in Table \ref{tab:cluster_stats}. For example, we can see that for the Julia Date SUT, after we merged all candidates found by any of the methods in any of the runs and clustered them we found a total of 11 clusters. In a 30 second run, LNS covered 4.9 +/- 0.3 of them and covered one cluster that was not covered by BCS (in a 30 second run), while in a 600 second run, BCS covered 7.5 +/- 0.8 clusters and covered 6 that were not covered by LNS (in a 600 second run). %FD check! fd: slight adjustments of wording. ok now.

LNS shows consistent but modest cluster coverage growth with increasing running time. In other words, boundary candidates found using LNS cover, on average, one or two more clusters when increasing the execution time from 30 seconds to 10 minutes. In contrast, BCS shows more cluster coverage improvement over time, where five additional clusters were covered when searching for 10 minutes, both for BMI Classification and Julia Date. It shows no such growth for bytecount or for BMI but it has also ``saturated'' for these SUTs already after 30 seconds, i.e. it has covered the total number of clusters found already after 30 seconds and thus have no potential for further growth.

\begin{table}
    \centering
    %\small
    \caption{Statistics over the potential boundary candidate clusters covered by both BCS and LNS. We also show the number of clusters that are uniquely covered by each approach, for each execution time setting. The Total Clusters column lists the total number of clusters found by the summarization method when run on all candidates found by any method in any run.}
    \label{tab:cluster_stats}
    \begin{tabular}{@{}llrrrrr@{}}
    \toprule
    \textbf{SUT} & \textbf{Strategy} & \multicolumn{1}{l}{\textbf{Total}} & \multicolumn{2}{c}{\textbf{30 seconds}} & 
    \multicolumn{2}{c}{\textbf{600 seconds}}\\
    \cmidrule(lr){4-5} \cmidrule(lr){6-7}
       &    & \multicolumn{1}{l}{\textbf{Clusters}} & \textbf{\# Found} &  \textbf{\# Unique} & \textbf{\# Found} & \textbf{\# Unique} \\
\midrule
\verb|bytecount| & LNS & 8 & $5.1 \pm 0.6$ & 0 & $6.0 \pm 0.0$ & 0 \\ 
 & BCS & 8 & $8.0 \pm 0.0$ & 2 & $8.0 \pm 0.0$ & 2 \\
\midrule
BMI-class & LNS & 15 & $13.7 \pm 0.8$ & 2 & $15.0 \pm 0.0$ & 0 \\ 
& BCS & 15 & $9.0 \pm 0.7$ & 0 & $14.2 \pm 0.8$ & 0 \\
\midrule
BMI  & LNS & 5 & $5.0 \pm 0.0$ & 0 & $5.0 \pm 0.0$ & 0 \\
& BCS & 5 & $4.8 \pm 0.4$ & 0 & $5.0 \pm 0.0$ & 0 \\
\midrule
Julia Date  & LNS & 11 & $4.9 \pm 0.3$ & 1 & $5.0 \pm 0.0$ & 0 \\
& BCS & 11 & $2.7 \pm 1.1$ & 1 & $7.5 \pm 0.8$ & 6 \\

    \bottomrule
    \end{tabular}
\end{table}

For many cases, BCS and LNS cover the same clusters, but there are some exceptions. For instance, only BCS found boundaries between the valid-error and error-error partitions for \verb|bytecount| --- clusters 7 and 8 in Table \ref{tab:bytecount_repr}. In contrast, and considering the 30-second search, only LNS identified candidates in BMI Classification that cover the transitions between (Underweight, Normal) and (Normal, Overweight) --- clusters 8 and 13 in Table \ref{tab:bmi_class_repr}. However, importantly, with increased execution time, BCS was the only strategy to find unique clusters (final column of Table~\ref{tab:cluster_stats}). Particularly, if we look at Julia Date (10 minutes), BCS covers 6 unique clusters --- including two clusters with ``valid'' outputs but unexpectedly long month strings. In RQ3, we further explain and compare these clusters for each SUT, and argue their importance.

The comparisons above highlight the trade-off between time and effectiveness of the exploration strategies. Overall, we see that LNS can be more effective in covering clusters in a short execution (BMI-class and Julia Data) but this is not always the case (bytecount). And with more execution time, BCS generally catches up to LNS (BMI-class) and often surpass it (bytecount and Julia Date) both on average and in the number of unique clusters that are uncovered. Clearly, attributes of the SUTs will affect cost-effectiveness, e.g., number of arguments in the input, (complexity of) specification, or the theoretical number of clusters that could be obtained to capture boundary behavior.

From Table~\ref{tab:cluster_stats}, we also note that the standard deviations are typically low so the method is overall robust to random variations during the search. Still, we do note that the best method for Julia Date (BCS) only finds 7-8 of the total of 11 clusters. This is not so for the other three SUTs where it tends to find all of the clusters in a 600 second run.

\vspace{0.4cm}
\begin{mdframed}[style=style1]
\textit{Key findings (RQ2):}
The identified boundary candidates cover a diverse range of boundary behaviors. While Local Neighbor Sampling (LNS) can sometimes find more clusters in a short (30 second) run, Boundary Crossing Search (BCS) catches up and often finds both more diverse and more unique candidates in longer runs. The framework is robust to random variations during the search and the number of unique behaviors found is mainly a function of the execution time and the characteristics of the SUT.
\end{mdframed}

\subsection{RQ3 - Identifying relevant boundaries}

None of the investigated SUTs have a formal specification to which we can compare the actual behavior of the implementations as highlighted by the identified boundary candidates. Formally, we can thus not judge if any of the identified boundary candidate pairs indicate real faults. 

Also, in practice, even if there was a formal specification it might not be fully correct or at least not complete, i.e. there might be situations\slash inputs for which it doesn't fully specify the expected behavior. Human judgement would then be needed to decide what, if anything, would need to be updated or changed in response to unexpected behavior uncovered during testing. In industry it is more common with informal specifications consisting of requirements in natural language, which can further exacerbate these problems. However, a relative benefit of black-box, automated boundary exploration with the techniques proposed here is that they can potentially help identify several of these problems be them in the specification (incompleteness or even incorrectness), the implementation (bugs), and\slash or both. And even if no problems are identified the techniques we propose can help strengthen the test suite.

Below, we go through each SUT, in turn, and manually analyse the boundary pairs identified and if they actually did uncover relevant (expected as well as unexpected) behavior or even indicate actual faults. We used the clusters identified by the summarization process (see RQ2 above) as the starting point. For clusters of a size smaller than 50 boundary pairs we went through all of them. For larger clusters we randomly sampled pairs, stratified by the total size of the outputs and analysed them, in turn, from smaller sized sub-groups to larger ones until saturation, i.e. looking at least at 50 pairs and going on further until no new, interesting or unexpected behavior was found. For additional detail we calculated also program derivative values using the Jaccard (based on 2-grams) function as output distance and checked all top-10 ranked pairs, per cluster. In the following, we highlight the key findings per cluster and SUT.

To support our reporting on the manual analysis we extracted tables with cluster representatives (see Tables \ref{tab:bytecount_repr} - \ref{tab:bmi_repr}). Unfortunately, for brevity, some of the tables' entries had to be shortened. The original values and details can be found as supplementary material \footnote{\url{https://github.com/feldob/repro_autobva/tree/master/clusterings}}. Since the answers to RQ1 and RQ2 above indicated that BCS sometimes was more effective (even if not always as efficient as LNS) the tables has a column indicating how many of the total candidates per cluster that was found by BCS.

\subsubsection{Bytecount}

Table \ref{tab:bytecount_repr} contains the representatives for the clusters identified for \verb|bytecount|. All 6 members of cluster 4 for \verb|bytecount| are the very natural and expected boundaries where the output string \textit{suffix} changes from a lower value to the next, e.g. from the smallest input pair of the cluster (999, 1000) with outputs (``999B'', ``1.0 kB'') to the largest pair (999949999999999999, 999950000000000000) with outputs (``999.9 PB'', ``1.0 EB''). While the behavior is not unexpected it is important that also these expected boundaries are identified. A tester can then more easily verify that the implementation actually corresponds to what is expected.

\begin{table}
    \centering
    \footnotesize
    \caption{Representative candidates for each cluster for bytecount including search coverage for BCS. Rows marked with an asterisk indicate clusters that are uniquely covered by BCS in a 600 seconds search.}
    \label{tab:bytecount_repr}
    \begin{tabularx}{\linewidth}{lXrrrrrr}
    \toprule
\textbf{ID} & \textbf{Validity Group} & \textbf{Input 1} & \textbf{Output 1} & \textbf{Input 2} & \textbf{Output 2} & \textbf{Cluster size} & \textbf{BCS found} \\
\midrule
1 & VV & -1 & -1B & 0 & 0B & 3 & 3\\
2 & VV & -10 & -10B & -9 & -9B & 33 & 33\\
3 & VV & 9950 & 9.9 kB & 9951 & 10.0 kB & 6 & 6\\ % Cluster 5 in the replication package!
4 & VV & 999 & 999B & 1000 & 1.0 kB & 6 & 6\\
5 & VV & 99949 & 99.9 kB & 99950 & 100.0 kB & 7 & 7\\ % Cluster 3 in the replication package!
6 & VV & false & falseB & true & trueB & 1 & 1\\
7$^*$ & VE & 99...56 & 1000.0 EB & 99...57 & BoundsError("kMGTPE", 7) & 1 & 1\\
8$^*$ & EE & 99...16 & BoundsError("kMGTPE", 9) & 99...17 & BoundsError("kMGTPE", 10) & 1 & 1\\
\bottomrule
\end{tabularx}
\end{table}

% Clusters 3, 5
The six members of cluster 3 has a similar pattern to the ones in cluster 4 but here the \textit{transition is within} each output string suffix category for the transitions from $9.9$ to $10.0$, e.g. the input pair (99949999, 99950000) with outputs (``99.9 MB'', ``100.0 MB''). Since the outputs in such pairs differ in length our output distance function detects them. Cluster 5 has the same six transitions but between $99.9$ and $100.0$, but also one extra boundary pair for the exabyte suffix (``EB''). Since this is the last suffix class and thus does not switch over to the next suffix at the value of ``1000.0 EB''. Since it is not obvious what the behavior at $1000.0 EB$ should be, not all specifications might cover it, and thus it would be important to test and check against expectations.

% Clusters 1
Cluster 1 contains three candidates all \textit{within} the ``B'' byte suffix group, covering the transitions from ``-1B'' to ``0B'', from ``9B'' to ``10B'', and from ``99B'' to ``100B''. While the transition from zero to negative one seems like a natural boundary, one could argue that the other two boundaries are less fundamental, and are an artefact of our specific choice of output distance function (string distance, here detecting the difference in lengths between "9" and "10" etc.). But the extra cost for a tester to verify that they are there seem slight. In the general case, there is, of course, a cost involved in having to screen very many candidate pairs. The transition from zero to negative inputs, however, should prompt a tester to consider if this should really be allowed (in the specification) or not.

The 33 members of cluster 2 are of more questionable relevance as they are all the transitions from ``-9B'' to ``-10B'', ``-99B'' to ``-100B'', and so on up for every output string length between 2 up to 36. An argument can be made that it is good for a tester to check if \textit{negative inputs} should even be allowed and, if so, how they are to be handled. But having more than a few such examples is probably not really adding extra insight, and the transition from 0 to -1 was already covered by the candidate in cluster 1 above.

The final valid-valid (VV) cluster (6) for \verb|bytecount| contains the single pair (false, true) with outputs (``falseB'', ``trueB''). This comes from the fact that in Julia the ``Bool'' type is a subtype of ``Integer'' and our tested Julia implementation of \verb|bytecount| only specifies that inputs should be integers; booleans are thus generated during the search and this pair is found. Again, it is not clear if this input type should actually be allowed but we argue it is important for a tester to know of this, implemented behavior to decide if it is good enough or needs to be addressed. Even if one decides to keep this functionality in the implementation it seems valuable to add it as a testcase to the test suite, at least as a kind of documentation.

There is a single valid-error (VE) cluster for bytecount (7) that has the single member (999999999999994822656, 999999999999994822657) where the first output is the string ``1000.0EB'' while the latter throws the exception $BoundsError("kMGTPE", 7)$. The Julia exception indicates that the implementation tried to access the string $"kMGTPE"$, of length 6, at position 7. Similarly there is a single error-error (EE) cluster (8) where the exception thrown changes from $BoundsError("kMGTPE", 9)$ to $BoundsError("kMGTPE", 10)$. Having found 3 inputs for which there are different kinds of BoundsErrors thrown it is then obvious that there will be other such transitions, i.e. between the errors accessing the string at position 7 and those at position 8 etc. Since our output distance only detects differences in length it doesn't identify the transition from 7 to 8 or 8 to 9 but picks up the transition from 9 to 10. This shows some of the trade-offs in the selection of the output distance function; while the one we have chosen here is very fast and does find a lot of relevant boundary pairs more fine-grained detection can be possible with more sensitive output distance functions. % For discussion: Future work should investigate this trade-off further.

\subsubsection{Julia Date}

Cluster number 4 for the Julia Date SUT, shown in Table~\ref{tab:JuliaDate}, contains a single boundary candidate pair which shows an unexpected switch in the outputs despite both being valid Dates. The pair also has among the largest program derivative values found overall (0.634). This candidate uses very large values for the year input parameter (757576862466481 and its successor), coupled with ``normal'' month and day values, but the outputs have no resemblance to the inputs and also switches the sign for the year in the Date outputs (outputs are 252522163911150-6028347736506391-02 and -252522163911150-12056695473012777-30, respectively). Even if such high values for the year parameter are not very likely in most use cases, we argue that it is still useful for a developer or tester to know about this unexpected behavior. They can then decide if and how to handle the case, i.e. to update either or both of the specification and the implementation or at least to document the behavior by adding a test case.

\begin{table}
    \footnotesize
    \centering
    \caption{Representative candidates for each cluster for Julia Date including search coverage for BCS. Rows marked with an asterisk indicate clusters that are uniquely covered by BCS in a 600 seconds search. Some input values and exception messages were abbreviated for brevity. `Err' refers to ArgumentErrors in Julia due to months (Mon) or days our of range (oor).}
    \begin{tabularx}{\linewidth}{lXrrrrXX}
    \toprule
\textbf{ID} & \textbf{Validity Group} & \textbf{Input 1} & \textbf{Output 1} & \textbf{Input 2} & \textbf{Output 2} & \textbf{Cluster size} & \textbf{BCS found} \\
\midrule
1$^*$ & VV & (-10000,2,3) & -10000-02-03 & (-9999,2,3) & -9999-02-03 & 8 & 8\\
2 & VV & (-1,9,3) & -0001-09-03 & (0,9,3) & 0000-09-03 & 115 & 38\\
3$^*$ & VV & (9999,5,9) & 9999-05-09 & (10000,5,9) & 10000-05-09 & 14 & 13\\
4$^*$ & VV & (75...81,2,21) & 25...50-60...91-02 & (75...82,2,21) & -25...50-12...77-30 & 1 & 1\\
5$^*$ & VV & (16...92,3,22) & 99...99-18...68-20 & (16...93,3,22) & 10...00-18...68-20 & 5 & 5\\
6 & VE & (0,2,0) & Err("Day: 0 oor (1:29)") & (0,2,1) & 0000-02-01 & 11019 & 1560\\
7 & VE & (330,5,0) & Err("Day: 0 oor (1:31)") & (330,5,1) & 0330-05-01 & 890 & 111\\
8 & EE & (-8,3,-1) & Err("Day: -1 oor (1:31)") & (-8,3,0) & Err("Day: 0 oor (1:31)") & 34465 & 6373\\
9 & EE & (0,0,92) & Err("Mon: 0 oor (1:12)") & (0,1,92) & Err("Day: 92 oor (1:31)") & 871 & 108\\
10$^*$ & EE & (0,4,99) & Err("Day: 99 oor (1:30)") & (0,4,100) & Err("Day: 100 oor (1:30)") & 8 & 7\\
11$^*$ & EE & (0,9...9,0) & Err("Mon: 9...9 oor (1:12)") & (0,1...0,0) & Err("Mon: 1...0 oor (1:12)") & 3 & 3\\
\bottomrule
\end{tabularx}
\label{tab:JuliaDate}
\end{table}

The pairs found in the valid-valid cluster 5 similarly all happen for large values of the year input parameter but differ from cluster 4 in that the output Dates are more similar to each other and typically only differ in one of the Date elements, e.g. year or month. Correspondingly the PD values are much lower (smaller variation around 0.20). 

Cluster 1 contains pairs where all years are negative and switches from one order of magnitude (all nines, e.g. "-9999-02-03", to the next one followed by zeros, e.g. "-10000-02-03"). Since the PD output distance is output string length many such boundaries (for different number of nines) are found. While the outputs in this cluster have similarities to the ones in cluster 5 above, the latter does not have inputs that corresponds to the outputs. For cluster 1 the input years corresponds to the output years. Splitting these into two clusters thus makes sense.

The largest valid-valid cluster (2) contains many pairs that only differ in the month and day while the year always goes from -1 to 0.

Clusters 6 and 7 both have pairs where one input leads to an ArgumentError while the other input leads to a valid Date. Both of the clusters has errors that either complains about an invalid Day or and invalid Month. We couldn't identify a clear reason why these two clusters were not merged. Most likely it is just an artefact of the clustering method we used in the summarisation step.

The remaining clusters all raise exceptions for both inputs. Cluster 11 has only 3 pairs which are all complaining about invalid Month inputs, all of a Month transition from a number of nines to the next order of magnitude. Similarly, the pairs of cluster 10 all complain about invalid Day inputs, all being variations of nines and the next order of magnitude. This cluster is larger since there are more unique exception of this type. The error message depends on the month since different months have differing numbers of allowed day ranges. Cluster 9 then has pairs where one input leads to argument error for invalid month and the other one for invalid day. Cluster 8 then mainly has both inputs being invalid day values although some combined pairs (both invalid month and invalid day) are also in this one.

\subsubsection{BMI classification}

For the BMI classification SUT (Table \ref{tab:bmi_class_repr}) there are many clusters that show the boundaries between ``adjacent'' output classes, i.e. Underweight to Normal (clusters 6 and 13), Normal to Overweight (8 and 11), Overweight to Obese (9 and 10), and Obese to Severly Obese (7 and 12). There are two clusters for each such boundary and they differ only in the order of the outputs, i.e. cluster 6 has the Underweight output first while cluster 13 has Normal first etc. 

We can also note that these clusters are relatively large with the smallest one (cluster 10) containing 93 pairs up to the largest one (cluster 13) cotaining 871 pairs. This is natural since the formula for calculating BMI allows for many different actual inputs being right on the border between two adjacent output classes.

\begin{table}
    \centering
    \footnotesize
    \caption{Representative candidates for each cluster for BMI classification including search coverage for BCS.}
    \label{tab:bmi_class_repr}
    \begin{tabularx}{\linewidth}{lXrrrrrr}
    \toprule
\textbf{ID} & \textbf{Validity Group} & \textbf{Input 1} & \textbf{Output 1} & \textbf{Input 2} & \textbf{Output 2} & \textbf{Cluster size} & \textbf{BCS found} \\
\midrule
1  & VV & (1,0) & Underweight & (1,1) & Severely obese & 19 & 19\\
2  & VV & (21,1) & Normal & (21,2) & Severely obese & 5 & 5\\
3  & VV & (26,1) & Underweight & (26,2) & Obese & 5 & 5\\
4  & VV & (29,1) & Underweight & (29,2) & Overweight & 3 & 3\\
5  & VV & (29,2) & Overweight & (29,3) & Severely obese & 2 & 1\\
6  & VV & (101,18) & Underweight & (101,19) & Normal & 511 & 459\\
7  & VV & (101,30) & Obese & (101,31) & Severely obese & 394 & 358\\
8  & VV & (108,26) & Normal & (108,27) & Overweight & 125 & 40\\
9  & VV & (115,32) & Overweight & (115,33) & Obese & 139 & 49\\
10 & VV & (132,44) & Obese & (133,44) & Overweight & 93 & 28\\
11 & VV & (133,41) & Overweight & (134,41) & Normal & 103 & 38\\
12 & VV & (1015,3087) & Severely obese & (1016,3087) & Obese & 814 & 747\\
13 & VV & (10...88,18...37) & Normal & (10...89,18...37) & Underweight & 871 & 818\\
14 & VE & (-1,0) & DomErr("H or W negative...) & (0,0) & Severely obese & 35176 & 3809\\
15 & VE & (1,-1) & DomErr("H or W negative...) & (1,0) & Underweight & 35051 & 3991\\
\bottomrule
\end{tabularx}
\end{table}

In contrast to the ``natural'' boundaries above, clusters 1 to 5 all contain fewer boundary candidates (from 2 to 19) but all correspond to transitions between output classes that are unexpected. For example, cluster 1 contains extreme examples of inputs that are very close but where the output class jumps all the way from Underweight to Severly obese. We note that all of these clusters happen for very extreme input values and it is likely that we can address many of these problems by putting limits on the valid ranges of each of the inputs. However, it is important that our method was able to find transition not only between some of these non-adjacent output classes but for several combinations of them.

Finally, the method identified a large number of valid-error pairs at either end of the output class adjacency scale. Cluster 14 has pairs that go from Severly obese to an input Domain error where one of the values are negative, while cluster 15 has pairs that go from the Underweight output class to input domain errors. The size of these clusters are very large and it is not likely a tester would get much extra benefit from having so many candidate pairs. Future work can thus explore ways of focusing the search to avoid finding very many candidate pairs that ends up in the same cluster.

\subsubsection{BMI}

For the BMI SUT (Table \ref{tab:bmi_repr}) there are only five clusters identified with clusters 3 to 5 all having one input that leads to a domain error raised while the other output is either NaN (cluster 3), Infinity (4), or 0.0 (5). While cluster 3 is rare and only happens for two specific input pairs, the two other clusters are very large. This reflects the fact that there are many ways to create an infinite output (height of zero and weight is any value) or zero output (weight is zero and height is any value).

\begin{table}
    \centering
    \footnotesize
    \caption{Representative candidates for each cluster for BMI including search coverage for BCS.}
    \label{tab:bmi_repr}
    \begin{tabularx}{\linewidth}{lXrrrrrr}
    \toprule
\textbf{ID} & \textbf{Validity Group} & \textbf{Input 1} & \textbf{Output 1} & \textbf{Input 2} & \textbf{Output 2} & \textbf{Cluster size} & \textbf{BCS found} \\
\midrule
1 & VV & (0,93) & Inf & (1,93) & 930000.0 & 5753 & 299\\
2 & VV & (106,11) & 9.8 & (106,12) & 10.7 & 99113 & 11181\\
3 & VE & (-1,0) & DomErr("H or W negative...) & (0,0) & NaN & 2 & 2\\
4 & VE & (-1,1) & DomErr("H or W negative...) & (0,1) & Inf & 95046 & 6534\\
5 & VE & (1,-1) & DomErr("H or W negative...) & (1,0) & 0.0 & 38449 & 4494\\
\bottomrule
\end{tabularx}
\end{table}

The valid-valid cluster 1 have pairs where the first outputs infinity while the second input leads to a normal, floating point output. The largest cluster is the valid-valid cluster 2 which has pair with normal, floating point outputs that differ only in their length. Of course, there are many such transitions and our system identifies many of them, but it is not clear that a tester would be helped by some of them more than others. Sorting just by length and including a few such transitions will likely be enough.

\subsubsection{Summary for RQ3}

Taken together, our manual analysis of the identified clusters and their boundary candidates shows that the method we propose can reveal both expected and unexpected behavior for each of the tested SUTs. Using bytecount as an example, 21 expected boundaries were automatically identified, divided into three main groups:

\begin{enumerate}
    \item transitions between consecutive byte suffix partitions, e.g. "999.9 kB" to "1.0 MB" (6 candidates),
    \item transitions from zero to negative values, "0B" to "-1B" (1),
    \item transitions within same byte suffix partitions, e.g. "9B" to "10B", "9.9 MB" to "10.0 MB", and "99.9 GB" to "100.0 GB" (14).
\end{enumerate}

Of these we argue the first two groups (1 and 2) are expected and natural while a tester can decide if and, if so, how many from group 3 to include in the test suite. The method also identified four (4) boundaries for bytecount that we argue were unexpected:

\begin{enumerate}
    \item transition from "999.9 EB" to "1000.0 EB",
    \item transition between boolean inputs "falseB" to "trueB",
    \item transition from valid output, "1000.0 EB" for input 999999999999994822656, to an exception, \verb|BoundsError("kMGTPE", 7)| for input 999999999999994822657,
    \item transition from two different exceptions, \verb|BoundsError("kMGTPE", 9)| for input 999999999999990520104160854016 and \verb|BoundsError("kMGTPE", 9)| for input 999999999999990520104160854017.
\end{enumerate}

In hindsight, it is likely a tester can understand the reasons for these boundaries, but we argue that it is not obvious from just looking at the implementation or a specification that these boundaries are there. Even though the very simple output distance function we have used cannot detect the additional error-error boundaries we can understand to be there (between bounds error 7 and 9, for example) it would be relatively simple to find them with a more focused search once we know to look for them. This also points to future work investigating alternative output distance functions or even hybrid search approaches that applies multiple distance functions, for different purposes.

For bytecount, another 33 boundary candidates were identified that were also unexpected but where we judge it less likely that a tester would include them all in a testsuite. These are the transitions between different sizes of negative inputs, e.g. from "-9B" to "-10B" and so on. Potentially, a tester might want to sample some of them to ensure proper treatment by a refined implementation but since the transition from zero to negative one has already been found the additional value is relatively limited.

For the Julia Date SUT, several clusters where of more debatable value and in particular in the error-error group it is likely that a tester would only have selected some typical examples from the identified clusters. While there were some differences between the clusters they essentially just differed in whether the month or day inputs lead to an exception being thrown. We do note that the boundary transitions for invalid day inputs covered all of the different months (30 to 31 valid-error transition for June, 31 to 32 for August, 28 to 29 for February etc). However, the clustering was not enough to separate them out into individual groups which made the manual screening less efficient.

For BMI-class, most clusters contained relevant boundaries. While all the expected boundaries between consecutive, ordinal outputs (like Normal to Obese) where identified, the method also identified a large number of unexpected boundaries between non-consecutive output categories.

For BMI, with its numerical outputs, a much larger number of candidates where identified. Still, the summarization method successfully distilled them down to only 5 clusters, which made it relatively easy to manually screen them. Even if expected and relevant boundaries were found it is harder, in this case, to define judge if there are other boundaries that should be found in the large, valid-valid groups. It was not clear, in this case that the output distance function used is fine-grained enough to pick out important differences.

While all eight clusters for bytecount and BMI contained at least one boundary candidate that we argue a tester would like to look at this was not the case for all the other SUTs. For example, for BMI-class several clusters differed only in the order of the outputs. This should be refined in future work on the method. There were also cases where clusters seemed to have been unnecessarily split into multiple clusters for which we could not discern any clear pattern or semantic reason. Most likely this is an artefact of the clustering method used and the features we used as its input. Still, we argue that since the total number of clusters identified was relatively limited it would not be a major cost for testers to screen them all.

While the number of identified candidates for bytecount was low (58), the clustering for summarization clearly helped in identifying interesting boundary candidates. This was even more evident for the more complex SUTs where the total number of candidates identified was very large; summarization is thus a necessity and clustering is one helpful way to achieve it. Future work should investigate how to refine the summarization further to decrease the number of candidates a tester has to look at.

\vspace{0.4cm}
\begin{mdframed}[style=style1]
\textit{Key findings (RQ3):}
The AutoBVA method can successfully identify both expected and unexpected boundaries without using a specification or white-box information. While the value of identified candidates ultimately depends on the tester, the summarization via clustering helped focus the manual screening. Further refinement to the summarization method should be investigated to try to minimize the number of different clusters a tester has to manually inspect.
\end{mdframed}

\section{Discussion}

Our results show that relevant boundaries in the input space and behavior of programs can be identified automatically, without the need for a specification nor for white-box analysis or instrumentation. This is important since it can help make boundary value analysis and testing more automated and systematic. While these techniques for quality assurance and testing have been advocated for long and sometimes even been required by standards and certification bodies, prior work has relied, for effective results, on the creativity and experience of the tester performing them. 

By building on the vision from~\cite{feldtdobslaw2019,dobslaw2020boundary} and coupling their proposal to simple search algorithms, the system we propose here enables augmenting the testers performing boundary value analysis and testing by \textit{automatically identifying and presenting them with candidate input pairs that are more likely to reveal actual boundaries}.

Our experimental validation shows that for the investigated SUTs the system could identify a large number of boundary candidate pairs that also cover a diverse range of behaviors. These results were also robust over multiple executions, despite the stochastic nature of the algorithms used. Manual screening showed that many relevant (important) boundary candidates were automatically identified, both those that could be expected for the investigated SUTs, and relevant but unexpected ones, that we argue would have been harder for a tester to think of.

We investigated two different search strategies within our overall framework. The simpler one, Local Neighbor Search (LNS) is more directly similar to random testing (in automated testing) but with a local exploration around a randomly sampled point. It  identified more boundary candidates but, even if given a longer execution time, didn't find as diverse types of candidates as the other strategy. The latter, Boundary Crossing Search (BCS), was tailored specifically to the problem at hand, by first identifying inputs in two different input partitions and then ``honing'' in on the\slash any boundary between them. BCS needs more computational resources but consistently found as many or more diverse clusters of candidates than LNS.

Regardless of the search strategy used, for our system to be useful to actual testers, a critical step is how to group and summarize the set of candidates found. While this is not the main focus of this study, we show that basing the grouping on the type of outputs of the pair and then clustering them based on their within- and between-pair distances, can be helpful. However, our experiments also uncovered challenges in this approach that should be investigated in future work, i.e. how to avoid showing too many groups (clusters) as well as candidates to a tester.
%Done: To discuss 2: Just acknowledge that the summarisation part was really not explored in depth in this paper and will be in future work.

The key idea that our system builds upon is that of the program derivative, a way to quantify the rate of change for any program around one of its inputs. A key choice we made in this study was to use a very fast but exceedingly simple distance function for outputs. By simply comparing the length of the outputs in a pair, after stringifying them, we can only detect a difference that leads to differing lengths. Clearly, this will not always be enough, for example for functions where all outputs are of the same length. Given this major limitation, our results are encouraging; we can find relevant boundaries despite it. One reason is likely that if outputs differ in some way they also tend to differ also in their length. Another reason is likely that by using such a fast but coarse distance function we can explore larger parts of the search space. Even the fairly simple Jaccard string distance function, that would detect more fine-grained differences in outputs, would be at least an order of magnitude and possibly more slower. And more advanced methods like the compression distances would be orders of magnitude slower yet. Future work should look into the trade-off between fast but coarse and slower but more fine-grained distance functions. We do note that the system need not select only one distance function though; hybrid solutions could be tried that first gets a coarser view of the boundaries and then zooms in for further details in different sub-partitions.

While our results constitute evidence that the overall approach has value, it is methodologically difficult to judge how complete a set of boundaries the method can find. Similarly, there is no clear baseline to compare to given that traditionally boundary value analysis has been a manual activity that is heavily dependent on both the existence of a specification and the experience of the tester. Still, future work should perform user studies, both "competing" with testers doing purely manual analysis and "in tandem" with testers to evaluate the added value of the automated system.

Our manual analysis of boundary candidates indicated that there are different types of candidates and they differ in what type of action they are likely to lead to. Below we give some examples and discuss these actions in some more detail. Finally, we then conclude the Discussion section by discussing limitations and threats to validity.

\subsection{Boundary candidates types and tester actions}

Table \ref{tab:boundary_situations} show five different actions that a tester can typically take in relation to one or a set of similar boundary candidates. The columns to the right shows the different artefacts that are likely to be affected in each case: the specification, the implementation, or the test suite.

A \textit{Skip} action typically means that either the tester does not find the candidate(s) relevant\slash useful, or it is already handled in the correct way by the implementation, clear enough in the specification, as well as tested well enough. 
Even in the latter case, it is important that an automated testing method can identify also boundaries of this type; if it would frequently miss them a tester might lose confidence in the tool.

An \textit{ExtendTests} action would happen when the identified behavior is both correct and according to the specification but not yet well covered by the test suite. 
The \textit{Debug} action occurs when a fault is identified in the implementation, i.e. the specification describes what the correct behavior should be but the implementation does not comply.
The \textit{Clarify} situation is a consequence of a behavior that is intended but not explicitly specified. Finally, the \textit{Refine} action is a consequence of a boundary that is not desired but when the specification is either incorrect or incomplete. In this situation we both need to clarify the specification and debug\slash fix the implementation. For all three of the last actions the test suite will also typically be extended to ensure the same problem does not occur in the future.

\label{sec:discussion}
\begin{table}
    \centering
    \caption{Different actions a tester can take given one or a set of related automatically identified boundary candidates, and which artefacts that are likely to change for each action.}
    \label{tab:boundary_situations}
    \begin{tabular}{llll}
    \toprule
    \textbf{Action} & \textbf{Specification} & \textbf{Implementation} & \textbf{Test suite} \\
    \midrule
    Skip & --- & --- & --- \\
    ExtendTests & --- & --- &  X  \\
    Debug     & --- &  X  &  X  \\
    Clarify   &  X  & --- &  X  \\
    Refine    &  X  &  X  &  X  \\
    \bottomrule
    \end{tabular}
\end{table}

One type of boundary candidate we identified in manual analysis was the \textit{under-specified input range}, i.e. for inputs that are not clearly expected which leads to unexpected behavior. One example, for BMI-class, was the candidate with outputs that move from underweight to severely obese by a change in body-weight from 0 to 1 kg for a person with a height of 1 cm. Another example Julia Date, with a very large year input that lead to nonsensical outputs that showed no resemblance to the inputs, likely due to some kind of overflow in the internal representation of date values.

This type of boundary would typically lead to either a \textit{Debug} or a \textit{Refine} action, depending on whether an update of the specification is meaningful\slash necessary or not.

Other types of problems were identified in relation to the \verb|bytecount| SUT. 
We used this SUT since it is known as the most copied (Java) code snippet on StackOverflow and has several bugs as described in \cite{Lundblad2019}: a rounding error, correctness for large values over 64 bits, 
and the fact that negative inputs are not appropriately handled. The latter was clearly identified, as described in our results above, can be seen as an \textit{under-specified} problem and likely leads to a Refine action since it is not clear from the specification how to handle negative inputs. The correctness for large values is still a problem even though in Julia the Integer type includes also 128 bit integers and the BigInt type to handle arbitrary large integers. However, the problem manifests in the boundary candidates that lead to BoundsErrors since they try to access larger byte suffix categories also after exabytes ("EB"). A Refine or at least a Debug action is probably called for. The boundary candidates capturing the rounding error of \verb|bytecount|, e.g. kB to MB, were all identified in the searches (cluster 5 in Table \ref{tab:bytecount_repr}) and represents \textit{wrongly positioned boundaries}. While it is not clear if the rounding error can be directly spotted even given the boundary examples in the cluster, it is contained in all the boundaries of the cluster. A diligent tester is likely to investigate key boundaries in depth and the fact that they are all in the same cluster can help make this checking more likely. This will lead to a Debug action.

\subsection{Threats to validity}
\label{limitations}

Our analyses involve simple comparisons through descriptive statistics (RQ1 + RQ2) and a qualitative analysis of the clusters (RQ3). We mitigate conclusion validity threats in our quantitative comparison by focusing on how the sampling strategies complement each other rather than simply comparing which one is better.
By analysing the uniqueness of candidates and clusters we identified essential trade-offs.
For RQ3, we do not perform an exhaustive inspection of all candidates in larger clusters, hence there is a risk we miss important ones. We mitigate this risk by random sampling and iterative analysis until saturation in conclusions.

We operationalize the constructs of our proposed approach with a limited set of treatments and dependent variables. Consequently, the main construct validity threats are our choices of: (1) configuration (string length for PD, mutation operators, sampling methods, etc.), (2) SUTs with integer inputs, (3) measures of uniqueness, clustering, and coverage\slash inspection of clusters.
However, given the novelty of the approach we argue it would be very difficult to analyse the overall results if we used more advanced or just more choices for these contextual factors. Future work is needed to better understand performance implications as these factors are varied.

We use clustering as an attempt to summarise boundary candidates into a presentable subset for testers. 
The ``relevance\slash importance of a boundary'' is to a certain degree subjective, hence an optimal or perfect summarization is hardly attainable.
It should be mentioned that in a real-world situation, a single search run with sufficiently many boundary candidates should suffice to do both the clustering and extraction of summary - which will both be faster and simpler than running many times over and building an overall clustering model over the entire set of found candidates. However, the number of clusters found in each run was smaller than the total number for some SUTs, even for BCS, so the effect of execution time needs further study in the future.

We mitigate internal validity threats by doing pilot studies and testing our instrumentation. The screening studies helped us to identify feasible and consistent strategies for sampling candidates and clustering outputs, instead of going for arbitrary choices. 
Encouraging was also that our method revealed a fault in our own implementation of the BMI value function (Table \ref{tab:bmi_repr})\footnote{The implementation did not check for height or weight equals to zero, hence triggering a division by zero error which lead to a NaN output.} Regarding the verification of the search procedure and system itself, we use automated unit tests in Julia to mitigate faults in our implementation.\footnote{\url{https://docs.julialang.org/en/v1/stdlib/Test/\#Basic-Unit-Tests}}

Lastly, our results cannot be generalised beyond the scope of our experimental study, i.e., finding boundaries for unit-level SUTs that take integers as input. On one hand, there are various aspects in AutoBVA that indicate it might be generally applicable, such as the concept of the program derivative and basing on string distances rather than type-specific distance function. However, more general test data generation algorithms, mutation operators, and distance functions will need to be experimented with to increase the external validity.

\section{Conclusions}
\label{sec:conclusions}

While automated boundary value analysis and testing is often advocated or required for quality assurance, prior work has relied mainly on the creativity and experience of the testers that perform them. Boundary value analysis has been a manual activity. Automated solutions are needed to better support testers, increase efficiency, and to make boundary value analysis more systematic. However, existing proposals have relied either on formal specifications, development of additional models, or white-box analysis, limiting their applicability.

Here we have presented an automated and black-box approach to identify and summarize boundaries of programs. It is based on coupling a boundary quantification method to a search algorithm. We further proposed to use string length distance as a very fast but coarse-grained boundary quantifier, and proposed two different strategies to search for boundary candidates. A clustering algorithm was used to summarize the identified candidates in groups based on the similarity of their values.

We validated our approach on four SUTs with both single and multiple numbers as inputs. 
We quantitatively evaluated how many candidates were found by the two search strategies as well as their uniqueness and diversity.
A manual, qualitative analysis of the identified boundary candidates was also performed to better understand their usefulness.

We find that, even using one of the simplest possible boundary quantification metrics, large quantities of boundary candidates could be found by both strategies. While the simpler Local Neighbor Search found more and more unique candidates than the Boundary Crossing Search strategy, the latter found more unique groups, if given more search time. Even though our approach is stochastic it was robust over multiple, repeated executions.
The manual analysis showed that many both expected and unexpected boundaries were found. Both previously known bugs in the investigated SUTs as well as new ones introduced by us were identified by the system. Based on our findings we outlined a simple taxonomy of different actions the proposed system can prompt a tester to take; refining either a test suite, an implementation, or even the specification.

While our results are promising, future work will have to consider more SUTs that are both more complex and have non-numerical input parameters. It should also explore more elaborate search strategies that can search globally over the input space as well as use the already identified candidates and groups to avoid unnecessary, repeated work.

%Bibliography
\bibliographystyle{IEEEtran}
\bibliography{references}

\appendix
\section{Screening Study - Configuring the Exploration Strategy}
\label{strategy_screening}

The exploration process has shown to be greatly influenced by a number of parameters - in particular regarding the sampling strategy. A pre-study investigating two parameters was conducted, and is detailed below. The purpose was to reduce the complexity of the main experiments without sacrificing quality if possible. We here describe the experiments and the verdict.

\begin{figure}[b]
    \centering
    \includegraphics[width=\linewidth]{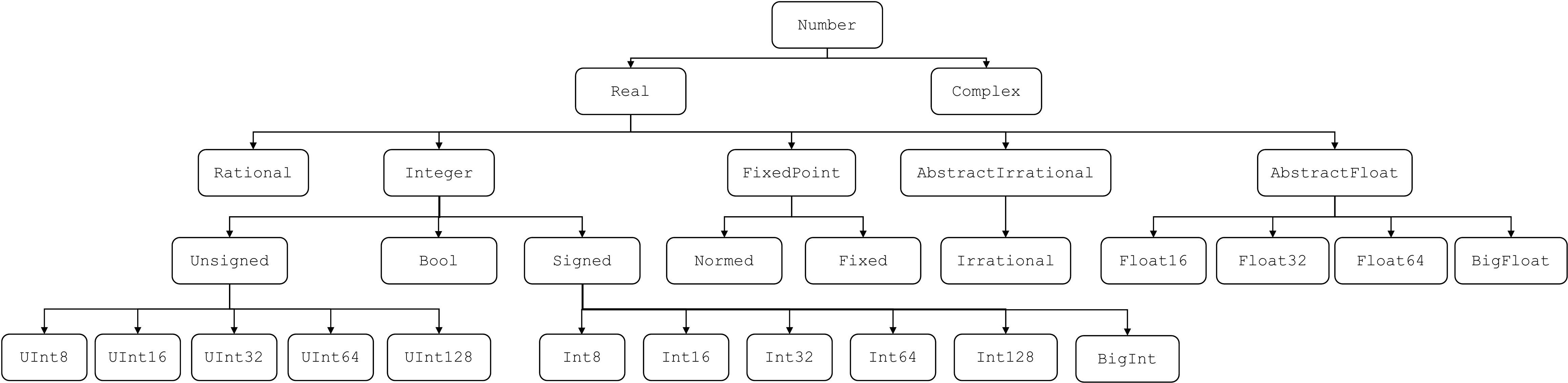}
    \caption{The type hierarchy for numbers in Julia showing the compatible types for Integer. Bool is an Integer.}
    \label{fig:julia_types}
\end{figure}

High level languages usually offer sampling based on a single concrete datatype (such as Int32 in Julia, representing Integers of 32 bits). When activating CTS, the sampling is instead done based on the compatible types per argument. For abstract data-types, the compatible types per argument can be derived from the type graph (see Julia's type graph for numbers in Fig. \ref{fig:julia_types}). In this study all input parameters are Integers. In Julia, Integers are an abstract type which cannot be sampled from. Thus we use Integers of base 128 in this study as per default. CTS must contain concrete types only, not abstract types, and a sampler for the type must be available. For instance, for Integer in Julia, the CTS in accordance with the graph is

$$CTS(Integer) = \{UInt8, UInt64, UInt32, UInt16, UInt128, Int8, Int64, Int32, Int16, Int128, BigInt, Bool\}$$

Further, the use of CTS is not limited to abstract datatypes such as Integer, but can be extended to concrete types. For Int16, and in accordance with the conversion rules of Julia, we may declare:

$$CTS(Int16) = \{UInt8, Int8, Int16, Bool\}$$

Sampling using CTS becomes a two-step process. For each argument, first, a compatible type is selected, and then the sampler for that type gets invoked. Two simple approaches to select the type for each sample is round-robin and uniformly at random. Without loss of generality, we use the latter approach in this study.

We here investigate the impact of CTS (activated, deactivated) and the sampling method (uniform, bituniform), both explained in \ref{sec:global_sampling}, on the efficacy of boundary exploration. We therefore apply AutoBD with LSN compared to BCS according to Section \ref{AutoBVA} on \verb|bytecount| for 30 and 60 seconds each for all possible configurations. As a boundariness measure, \emph{strlendist} of the outputs is applied. The results can be seen in Table \ref{tab:screening}.

\begin{table}
    \centering
        \caption{The results for the screening over the bytecount SUT.}
    \label{tab:screening}
    \begin{tabular}{clrlr}
    \toprule
\textbf{Time (s)} & \textbf{Algorithm} & \textbf{CTS} & \textbf{Sampling} & \textbf{\# Found} $(\mu \pm \sigma)$\\
\midrule
30 & LNS &  & uniform & $0.0 \pm 0.0$\\
30 & LNS & \checkmark & uniform & $8.6 \pm 0.9$\\
30 & LNS &  & bituniform & $7.0 \pm 0.0$\\
30 & LNS & \checkmark & bituniform & $9.8 \pm 0.8$\\
\midrule
30 & BCS &  & uniform & $0.0 \pm 0.0$\\
30 & BCS & \checkmark & uniform & $22.0 \pm 1.0$\\
30 & BCS &  & bituniform & $55.8 \pm 0.8$\\
30 & BCS & \checkmark & bituniform & $56.0 \pm 0.7$\\
\midrule
60 & LNS &  & uniform & $0.0 \pm 0.0$\\
60 & LNS & \checkmark & uniform & $9.8 \pm 0.4$\\
60 & LNS &  & bituniform & $7.8 \pm 1.3$\\
60 & LNS & \checkmark & bituniform & $10.4 \pm 0.9$\\
\midrule
60 & BCS &  & uniform & $0.0 \pm 0.0$\\
60 & BCS & \checkmark & uniform & $23.8 \pm 0.4$\\
60 & BCS &  & bituniform & $56.0 \pm 0.0$\\
60 & BCS & \checkmark & bituniform & $56.2 \pm 0.4$\\
\bottomrule
\end{tabular}
\end{table}

Note that regular uniform sampling \textit{without} CTS, what often is referred to as random search, almost never finds any boundary candidates. First when activating CTS boundary candidates can be identified. The greatest number of boundary candidates in all scenarios is though obtained when both CTS and bituniform sampling is activated. We therefore limit the study to configurations with CTS activated and bituniform sampling. The implied limitations in terms of generalizability are further discussed under \ref{sec:discussion}.

\section{Screening Study - Clustering for Summarization}
\label{sec:apx_clustering_screening}

The goal of our screening study is to identify a combination of features (U and WD) and distance measures (\emph{strlendist}, Jaccard, and Levenshtein) that yield a good model for doing k-means clustering of boundary candidates found by AutoBVA (as introduced and explained in section \ref{sec:summarization_detials}). Finding a good clustering is important to allow for an automated and consistent comparison of types of boundary candidates, since we cannot define equivalence partitions manually for the chosen SUTs. We also illustrate some examples of features extract from actual boundary candidates to clarify how the features and attributes represent the different types of boundaries.

In order to get a clustering of good fit we evaluate k in the range of 1--10, and select the one having the highest Silhouette score \cite{rousseeuw1987silhouettes} which offers an overview of the cohesion within each cluster and the separation between different clusters \cite{xiong2018clustering}. We use the default level of neighbors for orientation per data point of 15. We use Euclidean as distance metric between feature vectors, and 200 max iterations as per the interface default. Silhouette values vary between $+1$ and $-1$, where $+1$ indicates that the clusters are clearly distinguishable from each other; values of zero indicate that the clusters are relatively close to each other such that there is little significance in clustering the candidates. Lastly, $-1$ means that the distance between candidates within the cluster is greater than the distance between different clusters, hence indicating that the clustering performed is not appropriate and more distinct clusters are likely needed.

We repeated the runs 100 times per all combinations of features, totalling 64, and ranked the configurations according to the silhouette mean, while selecting the model of maximum score per configuration. The objective here was not to find perfect-fit models\slash clusterings but to identify models of good fit with high discriminatory power, because the more clusters that can be differentiated with high accuracy, the more diverse the individual boundary candidates to present to a tester.

The best models can be seen in Table \ref{tab:clusters_screening}. The selected feature-set was the one producing the largest number of clusters among the top five percentile silhouette scores for k-Means over $k \in {1,...,10}$, i.e. \emph{strlendist} (WD) and Jaccard (WD + U). It strikes a good balance between modeling quality and ability to discriminate for the \verb|bytecount| SUT that serves for the training due to its low computational complexity and straight forward separation of boundaries within the V domain. The clustering therefore uses the Jaccard metric for both WD and U. How/whether the feature space generalizes well is a question for future work. The final matrix with three features represented by four attributes over which all clusterings in this paper are conducted is therefore:

$$ M = \begin{bmatrix}
strlendist(WD)\\
Jaccard(WD)\\
Jaccard(U_1)\\
Jaccard(U_2)
\end{bmatrix}.$$

\begin{table}
    \centering
    \caption{The top 9 clustering configurations for bytecount with their respective average silhouette score and number of clusters (sorted by Silhouette Score).}
    \label{tab:clusters_screening}
    \begin{tabular}{lrr}
    \toprule
    \textbf{Configuration} & \textbf{Silhouette Score} & \textbf{Number of Clusters} \\
    \midrule
\textit{strlendist} (WD) + Jaccard (WD)     & 0.982 & 5\\
\textit{strlendist} (WD) + Jaccard (WD + U) & \textbf{0.942} & \textbf{6}\\
\textit{strlendist} (WD) + Jaccard (U)      & 0.938 & 3\\
Jaccard (WD + U)                            & 0.924 & 5\\
\textit{strlendist} (WD) + Levenshtein (U)  & 0.779 & 3\\
\textit{strlendist} (WD) + Jaccard (WD + U) + Levenshtein (WD) & 0.777 & 4\\
Levenshtein (WD) + Jaccard (U)              & 0.773 & 2\\
\textit{strlendist} (WD) + Levenshtein (WD) + Jaccard (U) & 0.771 & 3 \\
\textit{strlendist} (WD) + Jaccard (WD) + Levenshtein (U) & 0.760 & 4 \\
\bottomrule
    \end{tabular}
\end{table}

\end{document}